 \documentclass[preprint2]{aastex}

\begin{document}

\title{The Number Density of Old Passively-Evolving Galaxies at $z=1$ in the Subaru/XMM-Newton Deep Survey Field \altaffilmark{1}}

\author{
        Toru Yamada             \altaffilmark{2}
        Tadayuki Kodama         \altaffilmark{2,3}
        Masayuki Akiyama        \altaffilmark{4}
        Hisanori Furusawa       \altaffilmark{4}
        Ikuru Iwata             \altaffilmark{2,5}
        Masaru Kajisawa         \altaffilmark{2}
        Masanori Iye            \altaffilmark{2}
        Masami   Ouchi          \altaffilmark{6}
        Kazuhiro Sekiguchi      \altaffilmark{4}
        Kazuhiro Shimasaku      \altaffilmark{7}
        Chris Simpson           \altaffilmark{8}
        Ichi Tanaka             \altaffilmark{9}
        Michitoshi Yoshida      \altaffilmark{5}
}
\altaffiltext{1}{Based on data collected at Subaru Telescope, 
which is operated by the National Astronomical Observatory of Japan.}
\altaffiltext{2}{Optical and Infrared Astronomy Division, National Astronomical Observatory, Mitaka, Tokyo 181-8588, Japan.}
\altaffiltext{3}{European Southern Observatory, Karl-Scwarzschild-Str.2,
D-85748, Garching, Germany.}
\altaffiltext{4}{Subaru Telescope, National Astronomical Observatory of Japan, 650 N. A'ohoku Place, Hilo, HI 96720, USA.}
\altaffiltext{5}{Okayama Astrophysical Observatory, National Astronomical Observatory of Japan, Kamogata, Okayama, 719-0232, Japan.}
\altaffiltext{6}{Space Telescope Science Institute, 3700 San Martin Drive, Baltimore, MD 21218, U.S.A.}
\altaffiltext{7}{Department of Astronomy, School of Science, University of Tokyo, Tokyo 113-0033, Japan.}
\altaffiltext{8}{Department of Physics, University of Durham, South Road, Durham DH1 3LE, UK.}
\altaffiltext{9}{Astronomical Institute, Tohoku University, Aramaki, Aoba, Sendai 980-8578}

\begin{abstract}

 We obtained the number counts and the rest-frame $B$-band luminosity function of the color-selected old passively-evolving galaxies (OPEGs) at $z=1$ with very high statistical accuracy using a large and homogeneous sample of about 4000 such objects with $z' <25$ detected in the area of 1.03 deg$^2$ in the Subaru/XMM-Newton Deep Survey (SXDS) field. Our selection criteria are defined on the $i'-z'$ and $R-z'$ color-magnitude plane so that OPEGs at $z=0.9-1.1$ with formation redshift $z_{f}$=2-10 are properly sampled with minimum contamination by other population. The limiting magnitude corresponds to the luminosity of galaxies with $M_*+3$ at $z=0$. We made a pilot redshift observations for 99 OPEG candidates with $19 < z^\prime < 22$ and found that at least 78$\%$ (73/93) of the entire sample, or 95$\%$ (73/77) of these whose redshifts were obtained are indeed lie between $z=0.87$ and 1.12 and the most of their spectra show the continuum break and strong Ca H and K lines, indicating that these objects are indeed dominated by the old stellar populations. The relationship between the observed redshift and the color well follows the models used in defining the selection criteria in the consistent manner. We found that the surface number density of OPEGs varies 10-30$\%$ of the whole average from field to field even at a 30-arcmin scale and the poorest field, SXDS-South, has only 65$\%$ of the richest one, SXDS-East. We then compare our results with the luminosity functions of the color- or the morphologically-selected early type galaxies at $z=0$ taking the evolutionary factor into account and found that the number density of old passive galaxies with $\sim M_*$ magnitude at $z\sim 1$ averaged over the SXDS area is 40-60$\%$ of the equivalently red galaxies and 60-85$\%$ of the morphologically-selected E/S0 galaxies at $z=0$ depending on their luminosity evolution. It is revealed that more than half, but not all, of the present-day early-type galaxies had already been formed into quiescent passive galaxies at $z=1$.
\end{abstract}


\section{INTRODUCTION}

 Galaxies are expected to evolve passively after the major star formation ceased by consuming and/or losing a large fraction of their gas. Their photometric properties then change only due to the aging of the stars inside, which can be traced very well by simple stellar evolutionary synthesis models without suffering large effects by dust extinction or any consequence of on-going star-formation activity. Only merging or interaction with other galaxies may change their properties significantly.

 Arag\'on-Salamanca et al. (1993) first reported detection of systematic change of the colors in cluster galaxies at $z$=0.5-0.8 which are consistent with passive evolution models with formation redshift, $z_f$, larger than 2. This was followed by a large number of works studying passive-evolution properties in cluster and field galaxies at intermediate redshift (e.g., Yamada and Arimoto 1995; Stanford et al. 1997; Kodama et al. 1998; Zepf 1997). The expected characteristic colors of passive galaxies are also used to search for clusters of galaxies at high redshift (e.g., Dickinson et al. 1995; Yamada et al. 1997; Tanaka et al. 2000; Gladders et al. 2003). The survey for extremely-red objects (ERO) also sampled a number of old passively evolving galaxies (hereafter OPEGs) at high redshift (e.g., Cimatti et al. 2000;  Daddi et al. 2002; Miyazaki et al. 2003; Iye et al. 2003).

 Since galaxy colors are easier to be measured and modeled than other properties such as size, velocity dispersion, spectral line index, or morphology, it is useful to construct a sample of well-defined color-selected galaxies at each epoch, especially at high redshift, in order to firmly constrain galaxy formation and evolution scenarios. OPEGs are indeed suitable population for such a purpose since their colors are even more characteristic at high redshift and their photometric evolution is the simplest. 

 In this paper, we focus on the number density of OPEGs at high redshifts, since it provides a critical test for the build-up of such massive galaxies as predicted by the hierarchical formation of galaxies in the CDM Universe (e.g., Kauffmann \& Charlot 1999). There have been a number of attempts in this line measuring the number density of high-$z$ OPEGs and comparing it with local counterparts, such as COMBO-17 (Bell et al. 2004), K20 (Pozzetti et al. 2003), GDDS (Glazebrook et al. 2004), MUNICS (Saracco et al., 2005), and GOODS (Treu et al. 2005).  However, the results are often inconsistent, and nothing is conclusive yet, largely due to the limited spatial coverage probably.

 To remedy this situation, we here present a result from the Subaru/XMM-Newton Deep Survey (SXDS, Sekiguchi et al. 2005, in preparation) on the number density of the well defined sample of OPEGs at $z=1$. SXDS is designed to have unique combination of large area (1.2 deg$^2$) and depth ($B<27.5$, $R<27$, $i'<27$, and $z'<26$, all in AB system), which enables us to construct a sample of OPEGs down to a few magnitude below the characteristic luminosity, $L_*$, even at such high redshift. We analyze the distribution of the 4118 $z \sim 1$ OPEG candidates down to $z'=25$ selected from the total of 330,000 SXDS $z'$-selected galaxies in the SXDS $Ver1.0$ catalog (Furusawa et al. 2005, in preparation). The large volume coverage of SXDS is crucially important to suppress the uncertainties from field-to-field variation. Since strong clustering of OPEGs at high redshift is naturally expected in biased galaxy formation scenarios, large volume coverage is essential to sample a fair fraction of the universe, only with which we can discuss their averaged properties of galaxies. 

 With the current optical data-set, the highest redshift at which we can fully trace the OPEGs is $z\sim1$, since our longest wavelength filter $z'$-band can still sample the light from the wavelength longward of the rest-frame 4000\AA\ break. Also, at lower redshifts, OPEGs tend to be severely contaminated by blue foreground/background populations. Therefore, $z=1$ is the optimal choice of redshift to explore in the current study. We note, however, that in the very near future, the SXDS field is a subject of deep near-infrared (NIR) imaging in $JHK$ bands through the UKIDSS consortium (Warren 2003), which will then enable us to trace OPEGs toward much higher redshift up to $z \sim 3$ with a thorough photometric-redshift analysis.

 The model-dependent color selection must be calibrated by the detailed spectroscopic observations. For the purpose, we also conducted spectroscopic follow-on observations using the Subaru telescope. Although it is time consuming and our sample is still small, the results show the color selection criteria works well in the very consistent manner.

 The outline of the paper is as follows. In Sec. 2, we describe our color and magnitude criteria to select OPEGs and the effects of on-going minor star-formation activity, as well as the brief introduction to the SXDS data set. In Sec. 3 we describe our spectroscopic observations for the selected OPEG candidates. We then present the number counts of OPEGs and their variation across the SXDS fields in Sec. 4. The rest-frame $B$-band luminosity function is presented and discussed in Sec. 5.  For the related topics, the color properties of the general $z \sim 1$ galaxies (i.e., not only OPEGs but also bluer galaxies) in SXDS fields are discussed in our previous paper, Kodama et al. (2004). The clustering properties of the OPEGs will be also discussed in the future publication. Throughout the paper, we adopt a set of cosmological parameters as follows: the matter density $\Omega_m$=0.3, the cosmological constant $\Omega_\lambda$=0.7, and the Hubble constant $H_0$=70 $h_{70}$ km s$^{-1}$ Mpc$^{-1}$. 

\section{SAMPLE SELECTION}

\subsection{Subaru/XMM-Newton Deep Survey}

 Subaru/XMM-Newton Deep Survey (SXDS, Sekiguchi et al. 2005, in preparation) is a deep and wide-field optical/x-ray survey to observe a significant volume of the universe from intermediate to high redshift in order to study structure formation history of the universe with high statistical accuracy. The significant part of the area will be also observed in various wavelengths from ultra-violet to radio.  At optical wavelength, the field with 1.2 deg$^2$ area centered at ($\alpha$, $\delta$)=(02$^h$18$^m$, $-05^\circ$00$^{''}$) (J2000.0) was surveyed with Subaru Telescope (Iye et a. 2004) equipped with Suprime Cam (Miyazaki et al. 2002) in $B$, $R$, $i'$, and $z'$ band by late 2003. $B$- and $R$-band magnitudes are calibrated to those in Johnson-Cousins system and $i'$- and $z'$-band filters are to the Sloan Digital Sky Survey. The magnitude values are all measured in $AB$ system. The survey field consists of the five Suprime-Cam field of view, namely, SXDS-Center, -North, -South, -East and -West (hereafter SXDS-C, -N, -S, -E, and -W fields). The SXDS optical source catalog $Ver 1.0$ has been constructed by October, 2004. The detection limit of the shallowest part of the SXDS $Ver 1.0$ data set is $B=27.5$, $R=27.0$, $i^\prime=26.9$, $z^\prime=25.8$ ($5\sigma$ sky fluctuation with $2"$-diameter apertures). The seeing sizes of the composite images are 0.$^{''}$80-0.$^{''}$84. The properties of SXDS-$Ver 1.0$ discussed in this paper is summarized in Table 1. The detail of the data reduction and the photometry will be presented in Furusawa et al. (2005, in preparation). The colors used in this paper are those obtained with $2"$-diameter aperture after correcting the small ($\sim 0.05$ mag) aperture magnitude differences between each fields, which is due to the slightly different PSF shapes. We do not use the data within 300 pix of the edge to avoid possible artificial effects. We also avoid the area that is affected by the strong CCD blooming effect by the bright sources. The final effective area used in this paper is 1.025 deg$^2$.

\subsection{Photometric Sample of Old Passively-Evolving Galaxies}

 We selected a sample of old passively-evolving galaxies (OPEGs) in the SXDS field as follows. Fig.1 and 2 shows that expected colors and magnitude evolution of OPEGs with $5L_*$ and $0.1L_*$ (at $z=0$) formed at $z_f=10, 5, 3, 2,$ and 1.5 along the observed redshift, which are calculated using the Kodama and Arimoto (1997) evolutionary synthesis models. We then defined the criteria for $z=1$ OPEGs on the color-magnitude diagram; (i) $0.8 < i'-z' < 1.2$, and (ii) $-0.05 z' +3.01 < R-z' < -0.03 z' + 2.49$ to {\it pick up} OPEGs at $z=0.9-1.1$. The first criteria is to chose red galaxies with strong continuum break at around 8000 \AA . Note, however, this $i'-z'$ criteria alone is not sufficient to isolate the pure passively-evolving galaxies. We therefore adopted the second criterion which brackets the expected color-magnitude relation for OPEGs formed at $z_f=2-10$ and observed at $z=1$. Considering the photometric errors, we have further 0.05 mag margin at both red and blue side of the model loci.  

 Fig.3 shows the $z'$ versus $i'-z'$ color-magnitude diagram of the $z'$-selected galaxies in SXDS field. To avoid crowdedness in the figure, one-third of the whole objects, which are randomly selected by their positions are plotted. While we do not use any morphological criteria to select OPEG candidates above, the sources with the SExtractor stellarity index smaller than 0.9 are plotted by the smaller red dots in Fig.3 to demonstrate the galaxy color distribution. There is a red envelope in $i'-z'$ color as expected from spectral evolution of normal galaxies. 

 Fig. 4 shows the $z'$ versus $R-z'$ color-magnitude diagram for the objects which satisfy our criterion (i). The criterion (ii) is represented in the figure by the two tilted solid lines. While many of those $i'-z'$ color-selected red galaxies are distributed in the range defined by the criterion (ii), there is notable scatter in their $R-z'$ color. Among the 10655 objects with $0.8 < i'-z' < 1.2$ and $z' < 25$, $\approx 39\%$ (4118) satisfy the criterion (ii) while $\approx 25\%$ (2619) have the redder $R-z'$ color and $\approx 37\%$ (3918) have the bluer color. We selected the objects which satisfy criterion (i) and (ii) as the candidates of OPEGs at $z \sim 1$. Their sky distribution is shown in Fig.5. 

 The numbers of the detected OPEGs candidates are summarized in Table 2. Completeness and contamination for these color selection of OPEGs are discussed below and the calibration by the spectroscopic follow-up observation is presented in Section 3.

\subsubsection{Completeness in color selection}

  While our criteria are set so that we can pick up OPEGs at $z=0.9-1.1$ with $z_f=2-10$ completely, OPEGs with younger age or galaxies with some star-formation activity may run out from the color range. We already reported in our previous paper, Kodama et al. (2004), how the $i'-z'$ and $R-z'$ colors behave along the redshift for the models with different ratio of old and young (or 'bulge-to-disk') populations. From Fig.1 in this paper and Fig.1 in Kodama et al. (2004), one may see how our criteria work to reject galaxies at lower redshift or galaxies with young population.

  Here we consider more subtle cases, namely, the effect of a small fraction of young stellar components added to the colors of old passive galaxies. For example, if we add the spectrum with on-going star formation component to OPEG spectra with $z_f > 2$, $R-z'$ color becomes bluer rapidly. We check how much star-formation brings them out from our color range for OPEGs. We consider the two extreme cases, namely, constant continuous star formation and episodic starburst on a passively evolving model galaxy with $z_f=5$ with $L_*$ luminosity at $z=0$. First, we add the young stellar population with constant star-formation rate started at $z=5$ to this old population which has $i'-z'=1.05$ and $R-z'=1.98$ at $z=1$. We found that their $R-z'$ color becomes bluer than the limit ($R-z' \sim 1.7$) when we add the young population with more than 5$\%$ of the stellar mass while $i'-z'$ color becomes bluer than 0.8 only if we add more than 10$\%$. Second, we also consider the cases that we add young 'burst' (namely, episodic) stellar population. If we add the burst component with age of 0.2 Gyr with mass fraction of 2-3$\%$, $R-z'$ becomes bluer than the limit. If the age of the burst is as old as 0.5 Gyr, however, even adding the 10$\%$ mass of the young component does not remove the model out from our criteria. Only very young population seen in relatively short time scale brings them out from our criteria.

 On the other hand, Fig.2 shows that old and luminous galaxies at only slightly high redshift becomes redder than the limiting $R-z$ color while their $i'-z'$ color are still bluer than 1.2. So those galaxies with redder $R-z'$ color maybe galaxies with slightly higher than $z=1$, which implies sharp cut off in redshift to our sample.

\subsubsection{Contamination by dusty red objects}

 While previous search for extremely red objects (EROs) with $R-K > 3$ (AB) down to $K=22-24$ contains similar population of old passive galaxies at around $z=1$, it is known that a simple color cut also samples dusty red galaxies at intermediate or high redshift.
 For example, Miyazaki et al. (2003) reported that in their sample of EROs selected by the criteria $R-Ks > 3.35$ (AB) and $Ks < 22.1$ (AB), only 58$\%$ are the old galaxies at $z$=1-1.5. 
 With the Calzetti's extinction low, the blue objects at $z \sim 1-2$ with intrinsic spectra $f_{\nu}$=constant have colors $0.8 < i'-z' < 1.2$ when they suffer reddening with $1.0 < E(B-V) < 1.4$ (at $z=1$-2). Since objects with $E(B-V) > 1.25$ have $R-z' > 2.0$, some fraction of dusty red galaxies are thus rejected by our second criteria. Spectroscopic observations may tell more quantitative number for the contamination (see Sec.3)

 We would like to note that we do not use the term 'extremely red galaxies' for our sample since it is well in the range of the expected galaxy colors and not 'extreme' in any sense. We did not make any star/galaxy separation for the primary sample of the OPEG candidates analyzed below, either, although we made various tests including/excluding stellar sources selected by various threshold of stellarity index in the SXDS catalog.

\subsubsection{Completeness in observation near the limiting magnitude}

 We evaluated the incompleteness of our galaxy sample due to the photometric error along their apparent magnitude by using the model galaxies as follows. We first created the blank 'sky' images after removing the objects detected in the SXDS $Ver1.0$ catalog. The areas of the objects are substituted by the 'median' sky image constructed from three images at the different part of the same Suprime-Cam field. Then we rondomly put the model galaxies with the similar number density (of not only OPEGs but the entire $z'$-band selected objects) as observed on to these 'sky' images and counted the number of the objects recovered with the magnitude difference less than 0.25 mag and their fraction to the input ones. The model galaxies have de Vaucouleurs profile, assuming the most of the OPEGs are early-type galaxies, and their size are set to roughly follow the Kormendy relation. Fig.6 shows the completeness of such $z'$-selected galaxies in SXDS-C, -N, -S, -E, and -W fields and their average. One notes that the evaluated completeness are very similar for the five fields; at $z' = 25$, the completeness of the SXDS-S field with the shortest exposure time (11040 sec) is 67$\%$, that is not far from other fields with 70-73$\%$. The detection completeness is higher than 90$\%$ at $z'=23$, which corresponds to the OPEGs with $\approx M_*+1$. We note that the procedure adopted here may slightly (a few $\%$) overestimate the detection completeness since we did not fully took the effects of the source confusion and clustering into account. More comprehensive study of the detection completeness of the SXDS catalog will be discussed in Furusawa et al. (2005, in preparation).

\subsubsection{Photometric errors in color}

 It is possible that the sample is contaminated by the bluer or redder galaxies or becomes incomplete due to the systematic offset of the magnitude zero point as well as the random photometric errors at the faintest magnitude range since the number of the objects increases rapidly toward the fainter magnitude, especially for the blue ones.

 We first evaluate the possible effects of magnitude zero point errors as follows. We shifted the color criteria slightly blue or red, and examined how many the numbers of the selected galaxies change. If we shift the $R-z'$ colors alone 0.05 mag, either to the red or the blue side, the number of OPEG candidates decreases net 4$\%$ (14$\%$ and 15$\%$ drop off and 12$\%$ and 11$\%$ comes in to the sample when shifted to the red and the blue side, respectively) for the $z' < 25$ samples, which is only twice of the Poisson error of the sample. The sample seems stable for the effect. If we shift both of the colors $i'-z'$ and $R-z'$ to the same direction, however, the number difference becomes as large as 14$\%$; if we shift the criteria 0.05 mag bluer, the sample increase $10\%$, and if 0.05 mag redder, it decrease $14\%$. After all, the uncertainty of the number densities of the OPEGs evaluated by the color selection alone is likely to be 5-15$\%$ cannot be rejected. This fact reminds us of the importance of the spectroscopic calibration for the model-dependent color criteria. Our rather pilot study, presented in the paper, indicates that the relationship between the redshift and $i'-z'$ color follows the adopted model tracks very well (see Sec.3 and Fig.11 below). Larger sample of spectroscopic redshift or the robust photometric redshift for the entire sample after UKIDSS will be much helpful in isolating the red-sequence galaxies with rather model-independent empirical manner. 

 Next, we consider the effects of random photometric error of the bluer galaxies at the faint end. Typical errors in $R-z'$ and $i'-z'$ colors is 0.08 and 0.2 mag at $z=24.5$ and $z=25.5$, respectively. Assuming the Gaussian distribution of the photometric error with the dispersion of these values, we evaluated the possible contamination by calculating the objects coming in and dropped from the criteria. We ignore the effects due to the first stage, namely the $i'-z'$ color selection, and consider the effect on the $z'$ vs. $R-z'$ color-magnitude diagram. While the number of the OPEG candidates at $z'=24$-25 and 25-26 are 803 and 743, respectively, the net numbers of the contaminants are expected to be $\approx 0$ and 195, which yields the nominal contamination fraction $0\%$ and $26\%$, respectively. We here ignore the bias that the probability that the red galaxies are dropped out from the criteria is slightly larger than that blue galaxies come in due to the larger error in their color for the red galaxies. It is seen in Fig. 4 that the most of the contaminants are from the objects with $z' > 25.5$. Thus the OPEG sample seems not affected by the effect above $z'=25$, but it can be significantly affected by the contamination of the bluer objects at the fainter magnitude.

\begin{figure}
\figurenum{1}
\plotone{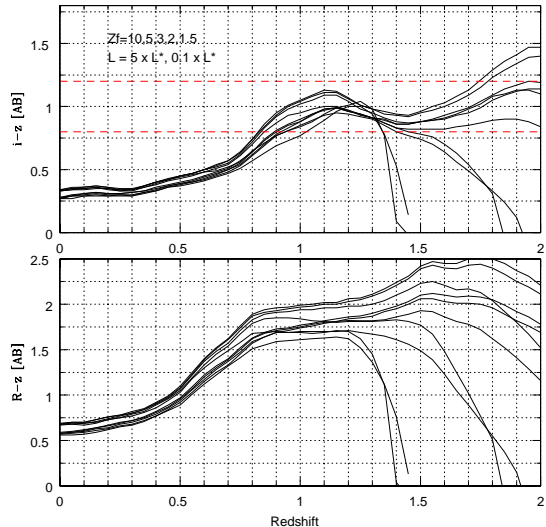}
\caption{$i'-z'$ and $R-z'$ colors of the model galaxies. The passive evolution cases with the formation epoch $z_F$=10, 5, 3, 2, and 1.5 (from red to blue) are shown for the galaxies with 5$L_*$ (thick lines) and 0.1$L_*$ luminosity at present. }\label{fig:fig1a}
\end{figure}

\begin{figure}
\figurenum{2}
\plotone{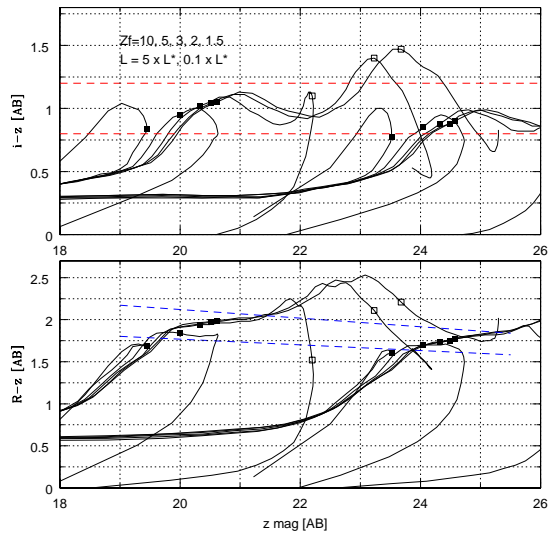}
\caption{Same as Fig.1 but the colors are plotted along the apparent magnitude. The filled and open squares show the redshift $z = 1$ and 2. The dashed lines are the color criteria for old passively-evolving galaxies adopted in this paper.}\label{fig:fig2}
\end{figure}

\begin{figure}
\figurenum{3}
\plotone{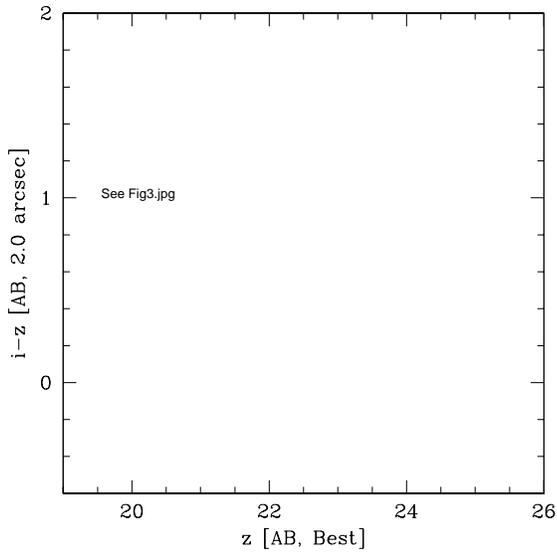}
\caption{$z'$ versus $i'-z'$ color-magnitude diagram of the $z'$-selected galaxies in SXDS field. To avoid crowdedness in the figure, one-third of the whole objects which are randomly selected by their positions are plotted. The red dots are those with stellarity index larger than 0.9. Dashed lines indicate the detection limits in $z'$ and $i'$ bands for the shallowest field. The solid lines show the adopted color criteria.}\label{fig:fig3}
\end{figure}

\begin{figure}
\figurenum{4}
\plotone{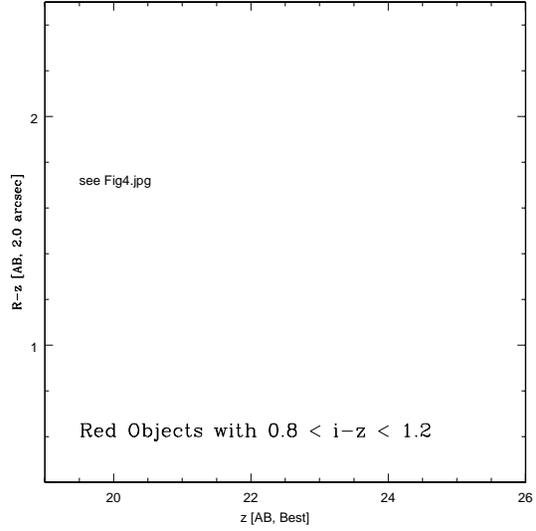}
\caption{$z'$ versus $R-z'$ color-magnitude diagram for the objects with $0.8 < i'-z' < 1.2$. Dashed lines indicate the nominal detection limits in $z'$ and $i'$ bands. Solid lines indicate the adopted color criteria for the color.}\label{fig:fig4}
\end{figure}

\begin{figure}
\figurenum{5}
\plotone{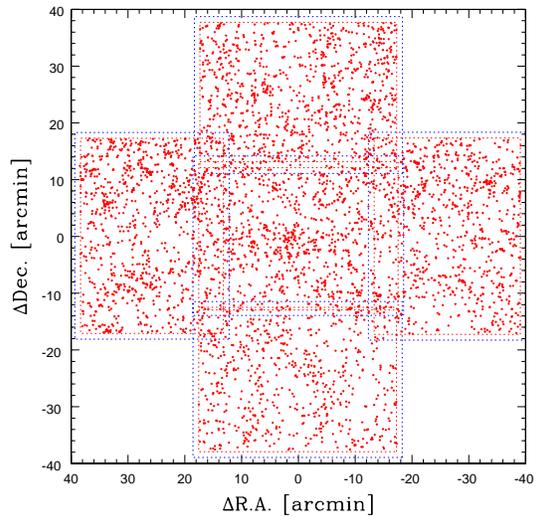}
\caption{The sky distribution of the selected OPEG candidates with $z'<25$ is shown.}\label{fig:fig5}
\end{figure}

\begin{figure}
\figurenum{6}
\plotone{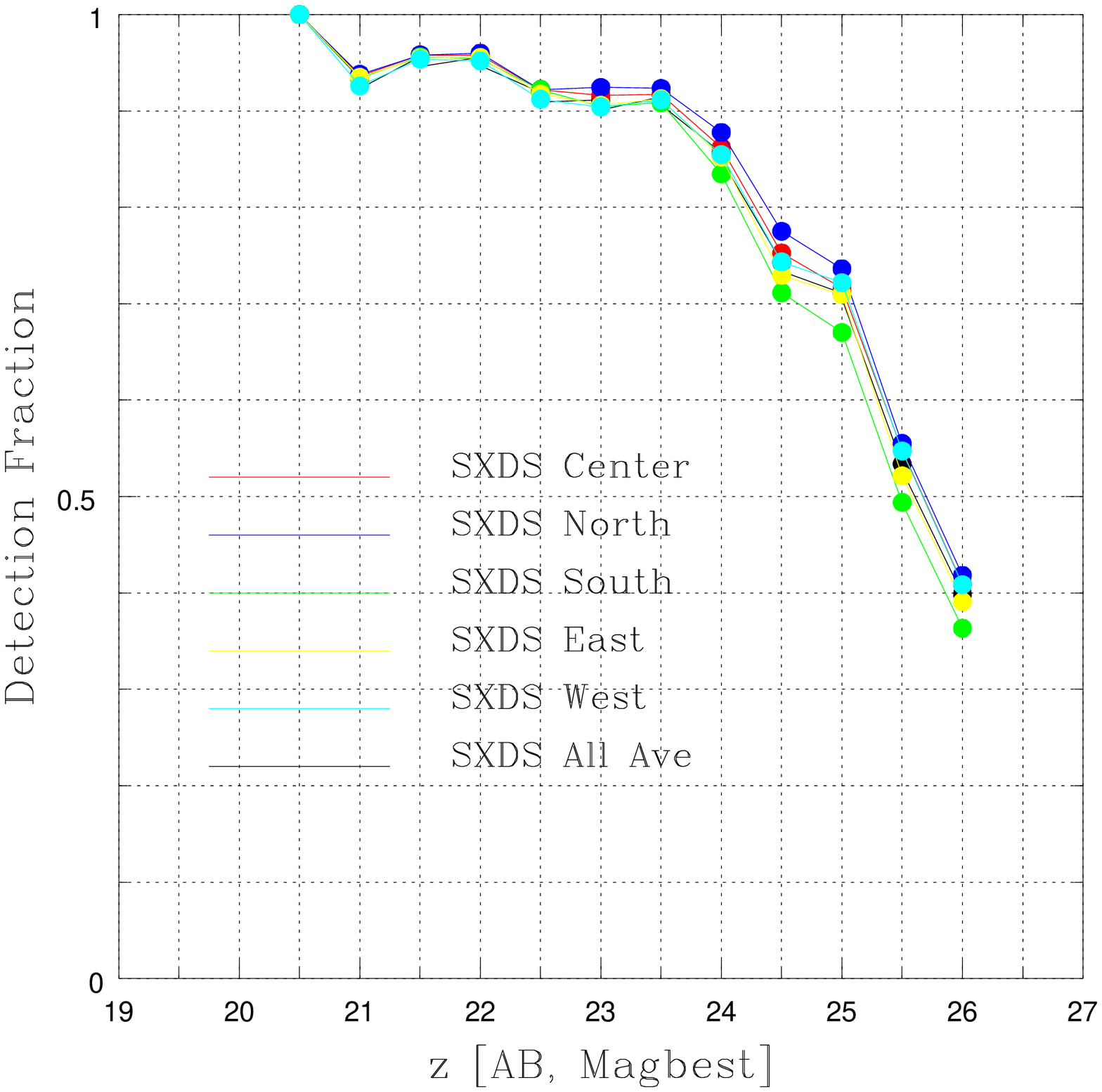}
\caption{Completeness of the detection of OPEG candidates on the $z'$ band images in SXDS-C, -N, -S, -E, -W fields evaluated by using the model galaxies and the observed background-noise images. De Vaucouleurs profile is assumed for the model galaxies.}\label{fig:fig6}
\end{figure}

\section{SPECTROSCOPIC SAMPLE}

 While we discussed on the possible contamination and incompleteness of our OPEG selection criteria in the last section, the true selection function can be only determined by spectroscopic observations. For the purpose, we conducted optical spectroscopy at a fraction of the SXDS field.

\subsection{Observations}

 In 2003 October and November, SXDS team conducted the follow-up spectroscopic observations of the optical/x-ray imaging surveys using Faint Object Camera And Spectrograph (FOCAS, Kashikawa et al. 2002) equipped with Subaru telescope. We observed more than 20 slit-aperture masks during the run. Many mask fields are chosen so that the interesting x-ray selected sources are to be observed but slitlets are also put on the OPEG candidates with $z' < 23$ in such fields. If there is enough room, we also put slitlets at OPEG candidates with $23 < z' < 24$ as well as the non-OPEGs with relatively small color difference. These target fields are therefore not homogeneously selected from the whole SXDS area but it is still unbiased for the red galaxies unless there is strong correlation between the x-ray sources and OPEGs at $z \sim 1$. While there are in fact some concentration of the x-ray sources at $z=1.1$ (Akiyama et al. 2005, in preparation), most of the other x-ray sources are at different redshift. We refer these masks as 'general field' masks for the OPEGs. Besides these masks, we also observed two masks in order to investigate the redshift distribution of the two 'high density' regions of red galaxies, C1 and C2, discussed in Kodama et al. (2004). In these two masks not only the red galaxies but also bluer galaxies are observed simultaneously.

 The sky distribution of the OPEG candidates, for which we observed their spectra, is shown in Fig.7 by the open and filled circles. In total, we observed 175 OPEG red galaxies with $19 < z' < 24$, of which 157 are in the 'general field' masks and other 18 are in the 'high-density region' masks. The 18 sources are located within $3'$ radius from the center of the regions C1 and C2 in Kodama et al (2004). In Fig.7 we also show the locations of the six high-density regions in Kodama et al. by the circle with 3$'$ radius. 

 We used the '150' grism except for the two masks we observed with the '300B'grism. Slit width is 0.8$"$ for all the cases which corresponds to the spectral resolution power of $R \sim 250$ or $\sim 30$ \AA\ at 8000 \AA\ for the 150 grism. The dispersion is 2.8 \AA\ per pixel. Exposure time is typically 3600 second. Unfortunately, these masks were not always observed  under the homogeneous condition; some exposures were obtained under cloudy/cirrus sky. 

 The data was reduced by using IRAF reduction package. After bias subtraction and flat fielding, each slitlet spectrum was extracted, and then wavelength and flux-sensitivity calibrated. The final wavelength determination accuracy is a few \AA. More detailed description of the FOCAS spectroscopy for SXDS sources is presented in Akiyama et al. (2005, in preparation). 

\subsection{Results for OPEGs}

 Table 3 summarizes the results for the OPEGs. At $z' < 22$, we observed 93 objects and the redshift of the 77 red galaxies (OPEG) were determined. 65 are in the 'general field' and 12 are in the 'high density' regions. At $22 < z' < 24$, our completeness in determining redshift becomes low and the redshift of 22 galaxies in the 'general field' were obtained. Table 4 lists the results of the redshift measurement.

 Fig. 8 shows the spectra for the 77 red galaxies with $z' < 22$ in order of the obtained redshift. Vertical lines show the wavelength of MgII $\lambda$2800 (red dashed line in Fig.8), FeI $\lambda$3581 (red), [OII] $\lambda$3727 (blue), MgI $\lambda$3835 (red), Ca H and K (green), H$\delta$ (cyan), H$\gamma$ (cyan), H$\beta$ (cyan) lines and G band (green).  For the most of the cases we see 4000 \AA\ continuum break and Ca HK lines clearly and found that their spectra are dominated by old stellar population. At the same time, [OII] emission line with the observed equivalent width $EW_{obs}>0$ \AA\ is detected in 29/77 objects. Only four galaxies have the rest-frame equivalent width larger than 10 \AA\ and the strongest one has 16 \AA . Many of these lines are thus weak compared with the emission-line galaxies even in the rich clusters (Poggianti et al. 1999) but they are thus not evolving passively in very strict sense. The objects with [OII] emission are plotted in Fig. 13 below and we found that many of them are among the $i'-z'$ bluer galaxies. There are also galaxies with weak but significant Balmer absorption lines (Fig.8). 

 Fig. 9 shows the histogram of the obtained redshift of the 77 red galaxies $z' < 22$ (solid line) as well as the in total of 99 red galaxies with $z' < 24$ (dotted line). The redshifts were determined by the lines in each spectrum.  The contribution of the 12 red galaxies in the high-density regions in Kodama et al. (2004) is also shown by the dashed line. We note that there is a strong concentration of the objects between $z=0.85$ and $z=1.1$ for galaxies with $z' < 22$. Indeed, 73 of the 77 objects with redshift ($95\%$) is lie between $z=0.87$ and 1.12.  The trend is also true for the fainter galaxies although the fraction of the galaxies at $z>1.1$ is higher. Sampling galaxies in high density region does not cause any difference (see below for the reality of the structure). Fig.10 presents the same result in a different way; the redshift values are plotted against the $z'$-band magnitude.

 In Fig.11, we plotted the observed $i'-z'$ color of the objects along the redshift. At $0.8 < z < 1.1$ there is strong positive correlation between the $i'-z'$ color and redshift. This is indeed expected for galaxies following the passive evolution. We plotted the passive-evolution models with $z_f$=2 and 5 and L=5L$_*$ and 0.1L$_*$ in the figure. Although the scatter exists, the data and the models agree quite well. From this figure, we strongly argue that these galaxies with redshift indeed obey passive evolution and their red $i'-z'$ color is due to the large 4000 \AA\ break of the old stellar population. Our color selection of OPEGs thus works very well.

 There is an obvious exception at $z>1.1$; a galaxy at $z \sim 1.4$ has much redder color than the OPEG models. This is SXDS J021754.65-045027.6 at $z=1.393$. Since HK lines and the 4000\AA\ break are redder than our spectral range, we cannot say much about the nature of this galaxy or probably a reddened AGN. 


 Of the 93 galaxies with $19 < z' < 22$ selected by our color criteria, we could not determine the redshift of 16 objects. While it is true S/N ratio of spectra is not enough for some objects, which is due to poor sky condition, many of the spectra of these objects are rather featureless to securely determine their redshift. In Fig.12, we plot one example of such red objects without any notable spectral feature to measure their redshift.

 We also obtained the spectra of non-OPEGs at $0.85 < z < 1.1$ in the same mask serendipitously. Fig. 13 shows the redhisft-$i'-z'$ distribution of these galaxies as well as those of the OPEGs (the filled ciecles). Objects with $EW_{obs}(OII) > 5$\AA\ are pltted with squares. Most of the bluer galaxies have the [OII] emission. This figure demonstrates how our selection criteria work to select out the passively-evolving galaxies. First, the OPEGs with [OII] emission have the relatively blue colors above the criteria. It is then natural to consider that a fraction of OPEGs or their progenitors drop out from our criteria due to a little more amount of on-going star formation. On the other hand, some galaxies have bluer $i'-z'$ color but no [OII] emission. They may be the objects without significant on-going star-formation but with younger stellar population, such as the $E+A$ galaxies seen in the intermediate-redshift clusters (e.g., Dressler et al. 1999), or the younger OPEGs with formation epoch less than $z=2$. The fraction of the [OII]-detected objects is higher at $z>1$ (19/36) than the lower redshift (10/41). This may be partially due to the selection effect since the colors of the OPRGs become redder as redshift increase and thus the rejection of the bluer galaxies are more sharp at the lower redshift. 

\subsection{Galaxies in the high-density regions}

 Fig.14 shows the redshift distribution of the galaxies in the high-density regions of red galaxies in Kodama et al. (2004). The sample is dominated by the objects whose $Ri'z'$ colors are consistent with the model galaxies at $z \sim 1$ with various ratio between old and young stellar population. Despite our speculation that there may be rich clusters of galaxies in these high-density regions, especially at C1, the redshift distribution does not show any strong single peak or concentration. C1 and C2 are likely to be a projection of a few groups. C1 has a rather discrete peak at $z=1.05$.

 We also note that this result does not affect our conclusion in Kodama et al. (2004) since these high density regions of red $z\sim1$ galaxies were selected in order to get the statistical field subtraction method work, and it does not really matter whether it is a single rich cluster at $z\sim1$ or a projection of a few groups at similar redshifts.

\section{NUMBER COUNTS}

 As described in Sec.2 and Sec.3, we constructed a sample of OPEG candidates at $z=1$ with $z_f=2-10$ by the carefully determined color and magnitude criteria. Spectroscopic results, especially for brighter galaxies, confirmed that the selection is highly efficient and at least 78$\%$ of the entire sample or 95$\%$ of those whose redshift were measured are indeed galaxies at $0.85 < z < 1.1$ whose spectra are dominated by old stellar population. If we add a small number of objects at $1.1 < z < 1.4$, the fraction of OPEG is more than 83$\%$ of the entire color-selected sample. From this fact, we may argue that our {\it photometric} sample is robust as it is dominated by OPEGs (we thus hereafter refer the color-selected objects as 'OPEGs' instead of 'OPEG candidates' for convenience).

  In this section, we then show the magnitude-number relation of these OPEGs and their field-to-field variation at large scale. The luminosity function after taking the selected redshift range into account is then discussed in the next section.

\subsection{The average number counts}

 Fig.15 shows the number counts of the whole $z'$-band selected galaxies and as well as OPEGs. For OPEGs, we also plotted the counts after correcting the incompleteness (Fig.6) evaluated for model galaxies with de Vaucouleurs light profile.
 For comparison, $z'$-band count obtained by Capak et al. (2004) with the same instrument and filter is plotted. Although they agree with reasonable level, there is still slight (a few to several $\%$) systematic difference  between the counts, which corresponds to the $\approx$ 0.05 mag zero-point difference. This may be due to the field-to-field variation or the different method in evaluating the total magnitude; Capak et al. (2004) used $3"$-diameter aperture magnitude and  aperture correction factor while we used the SExtractor MAG BEST values directly.

 While the $z'$-band number counts keep rising even at $z'=25$, the counts for OPEGs begin to flatten at $z' \sim 22$ and decrease toward the fainter magnitude even after the correction for the incompleteness.  This is not due to the underestimation of the incompleteness since, at $z' \sim 24$, OPEGs are in fact relatively more compact than the whole $z'$-selected ones and suffer less from detection incompleteness. The mode of FWHM values for OPEGs is  4.5 pix while that of the whole $z'$-selected objects is 4.9 pix.  Thus, while the number of the $z'$-selected sources which are more diffuse than OPEGs still increase, the number of OPEGs decrease along the magnitude.  For the cases of pure passive-evolution models, the L$_*$ galaxy at $z=0$ is to be $z' \approx 22$ at $z=1$. Therefore, if the selected red galaxies are dominated by OPEGs, the flattening at around $z'=22$ is what we expect for the case due to the shape of the galaxy luminosity function.

\subsection{Field-to-field variation at large scale}

 We then investigated how the counts vary from field to field at large scale. In Fig.16, we plot the number counts of OPEGs in each SXDS-C, -N, -S, -E, -W fields after the correction for the incompleteness. We found quite a large variation among them, especially in SXDS-S; at $z'=23$, the highest counts, of SXDS-N, is about twice as large as the lowest counts of SXDS-S. Note that the evaluated completeness at $z'=23$ is 90-92$\%$ in SXDS-N and -S, respectively. The $z'$-band detection limit is also comparable, 25.7 and 25.5 for SXDS-N and -S, respectively. Therefore, we believe that the difference of the number counts of OPEG candidates is not an artifact but the real nature of galaxies. We checked that the total counts of the $z'$-selected galaxies have the much less scatter distribution among the fields, at most 5$\%$ (Table 2).

 To confirm that it is not due to the inhomogeneous sample selection due to a systematic color difference among the fields, we also compare the $R-z'$ color distribution of the galaxies with $0.8 < i'-z' < 1.2$ in Fig.17. While the locations of the peaks of the color distribution of the five fields at around $R-z \approx 1.85$ agrees very well, a significant deficit of the galaxies in SXDS-S is seen at $1.6 < R-z < 2.0$, namely the color range of OPEGs. 

It is remarkable that the counts of OPEGs varies significantly even at this large scale. The each Suprime-Cam field of view has approximately $34'$ $\times$ $27'$, which corresponds to  46 $h_{70}^{-1}$ Mpc $\times$ 36 $h_{70}^{-1}$ Mpc in comoving scale at z=1. The nominal comoving volume at $0.85 < z < 1.1$ per field is $1 \times 10^{6}$ $h_{70}^{-3}$ Mpc$^3$. The distribution of the massive oldest galaxies appears inhomogeneous at 100 Mpc scale at $z=1$.

In Fig.16, not only the heights but also the shapes of the number counts show variation among the fields, in particular, the counts in SXDS-S show a rather flat slope at the faint end, while the counts in SXDS-N show a significant peak at around $z'=23$ and a negative slope at the faint end. The flat slope in seen in SXDS-S may be caused by relatively larger contamination, since the overall number density of OPEG is the lowest in this field.  The blue galaxies, which are abundant, tend to be scattered into our color cut due to their photometric errors. Also, the background galaxies that have similar colors as those of the $z=1$ OPEGs inevitably come into our sample especially at the faint end. We repeated the similar procedure in Sec.2.2.4 in the SXDS-S region and found that the contamination due to the random photometric error is larger in this field compared to other fields, but is still only $2\%$ for the objects at $z'=24$-25. As for the contamination by the background galaxies, since our spectroscopic sample is nearly complete only down to $z' \sim 22$, the true redshift distribution of the sample at the faint end is unknown. If the background contamination in the SXDS-S field is as large as $\approx 50\%$ at $z' = 24$-25, the slope of the faint end would become negative and all the fields would have consistent slopes after subtracting the background.

We note that Kodama et al. (2004) reported a much stronger deficit (a factor of 2-3 from $z'=22$ to 24) of red faint galaxies at $z\sim1$ in the same SXDS field, even though they applied a similar colour selection to ours based on $Riz'$. As shown in our Fig.14, the {\it averaged} number counts of the OPEG candidates over the entire SXDS field does not show such a prominent deficit towards the faint-end, although the slope is clearly negative. It should be reminded, however, that Kodama et al.'s (2004) sample of red galaxies are different from that in this paper in two ways: First of all, their sample were constructed in the selected high density regions of red $z\sim1$ galaxies, rather than the entire SXDS field used here. Secondly, they have applied a statistical subtraction of foreground/background contamination using the mean density of such red galaxies outside the high density regions. 
 To discuss more on this interesting issue on the deficit of faint red galaxies (which is often referred to as "down-sizing"), we badly need more accurate photometric redshift information on these faint galaxies by adding deep NIR imaging which will be provided in the near future through the UKIDSS collaborations.

\clearpage


\begin{figure}
\figurenum{7}
\plotone{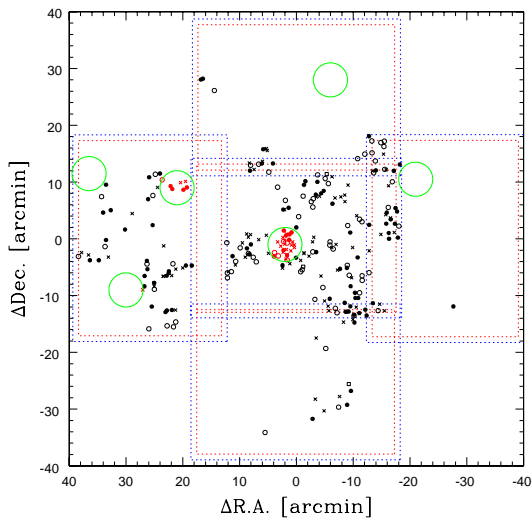}
\caption{The sky distribution of the targets in the spectroscopic observations. The circles are the OPEG candidates and the squares are other objects.  Filled symbols are those for which redshift were obtained. Large circles show the high-density regions of red galaxies studied in Kodama et al. (2004). }\label{fig:fig7}
\end{figure}

\begin{figure}
\figurenum{8}
\plotone{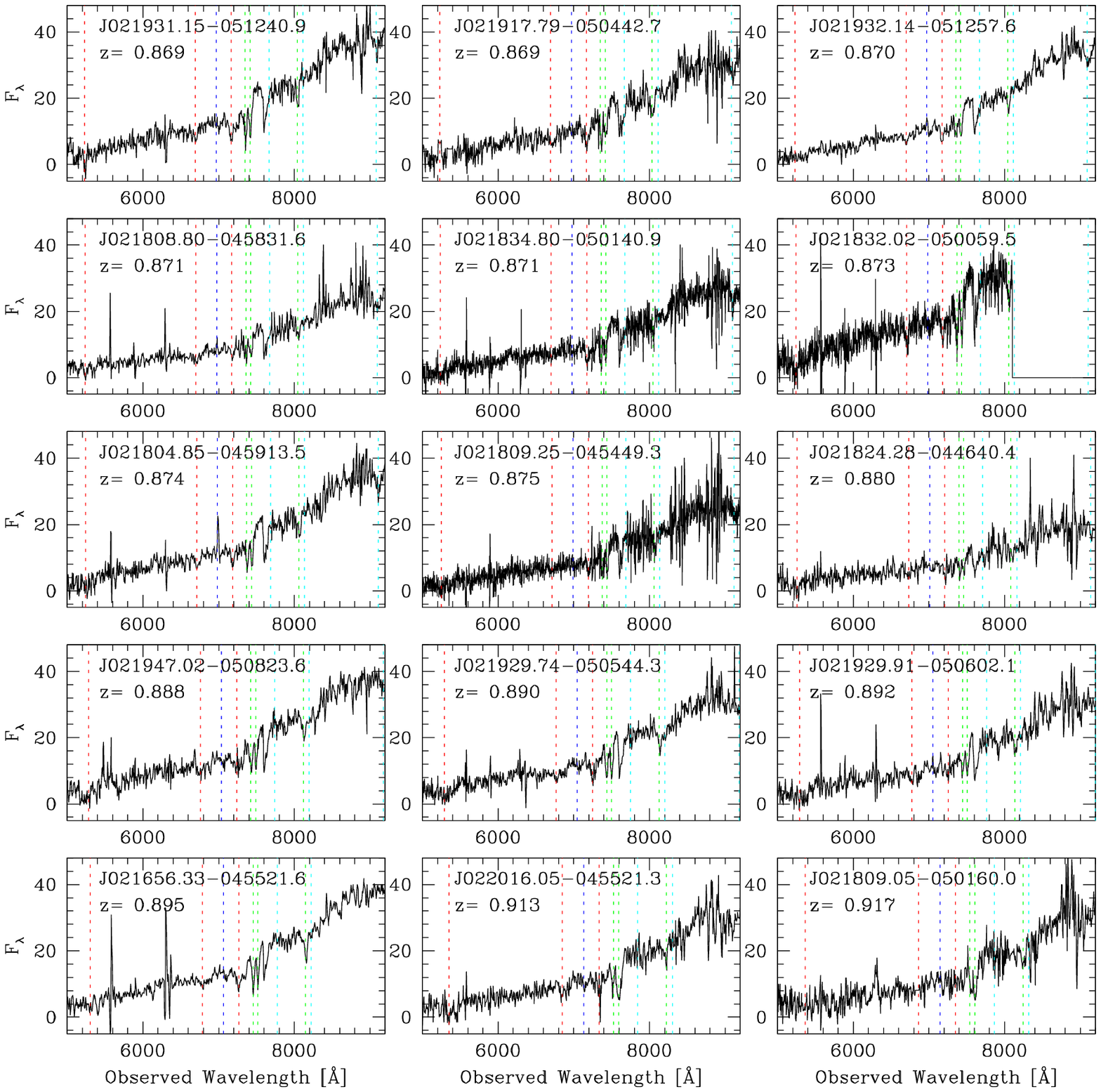}
\caption{The obtained spectra of OPEG candidates with $z' < 22$. Vertical dashed lines from left to right show the wavelength of the features of MgII $\lambda$2800 (red in the on-line edition), FeI $\lambda$3581 (red), [OII] $\lambda$3727 (blue), MgI $\lambda$3835 (red), Ca H and K (green), H$\delta$ (cyan), H$\gamma$ (cyan),  G band (green), and  H$\beta$ (cyan). }\label{fig:fig8a}
\end{figure}

\begin{figure}
\figurenum{8}
\plotone{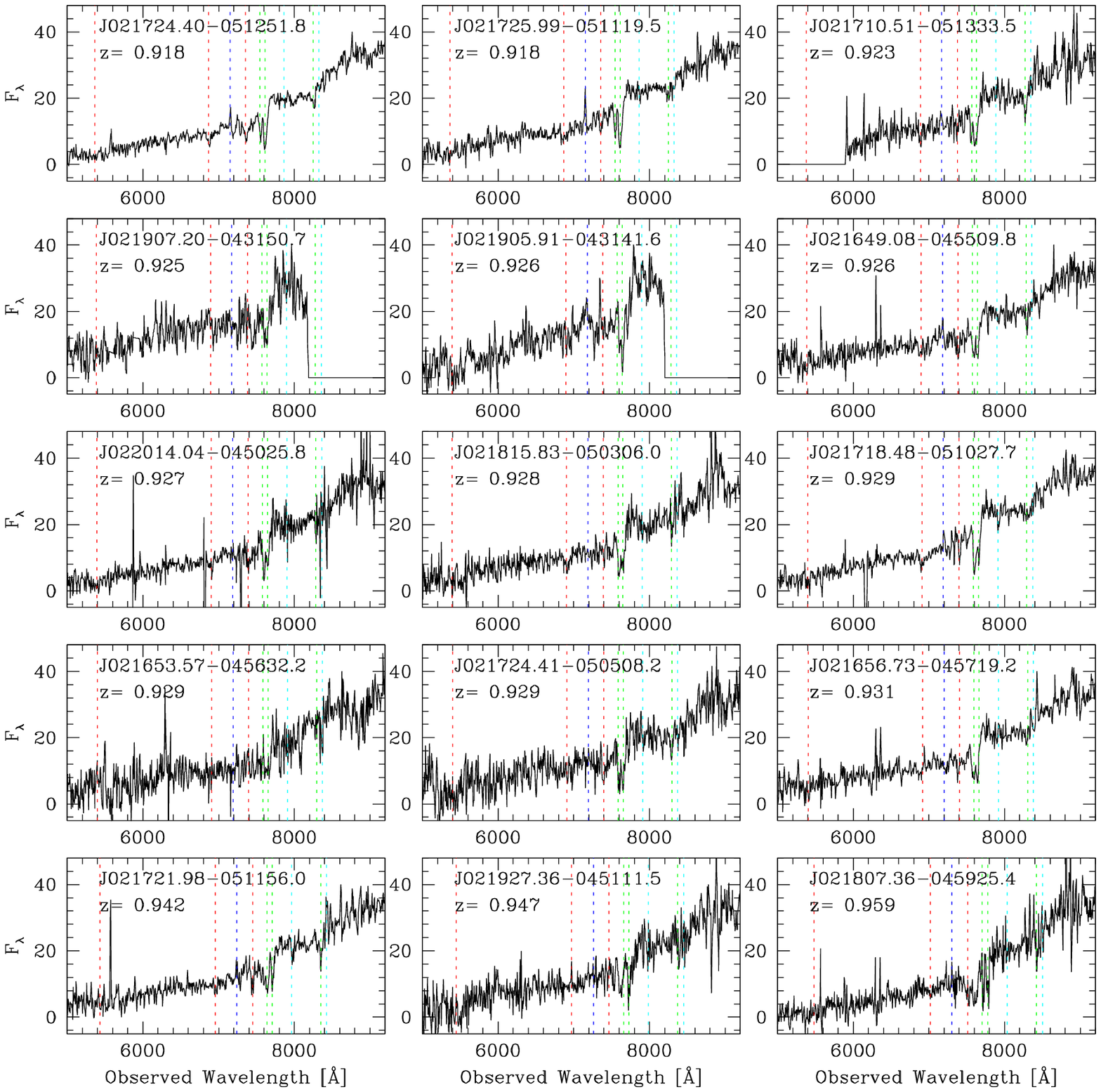}
\caption{Continued.}\label{fig:fig8b}
\end{figure}

\begin{figure}
\figurenum{8}
\plotone{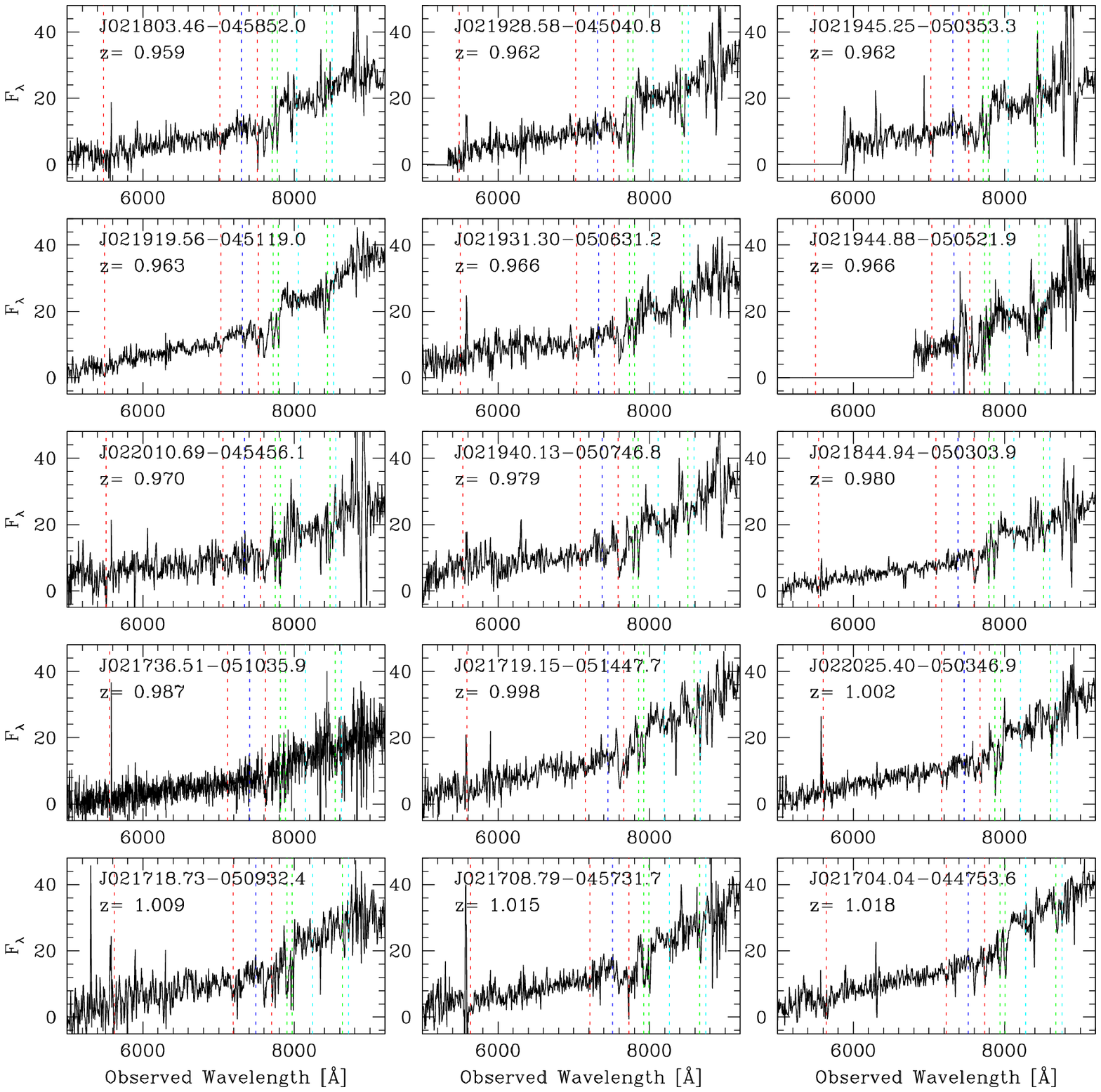}
\caption{Continued.}\label{fig:fig8c}
\end{figure}

\begin{figure}
\figurenum{8}
\plotone{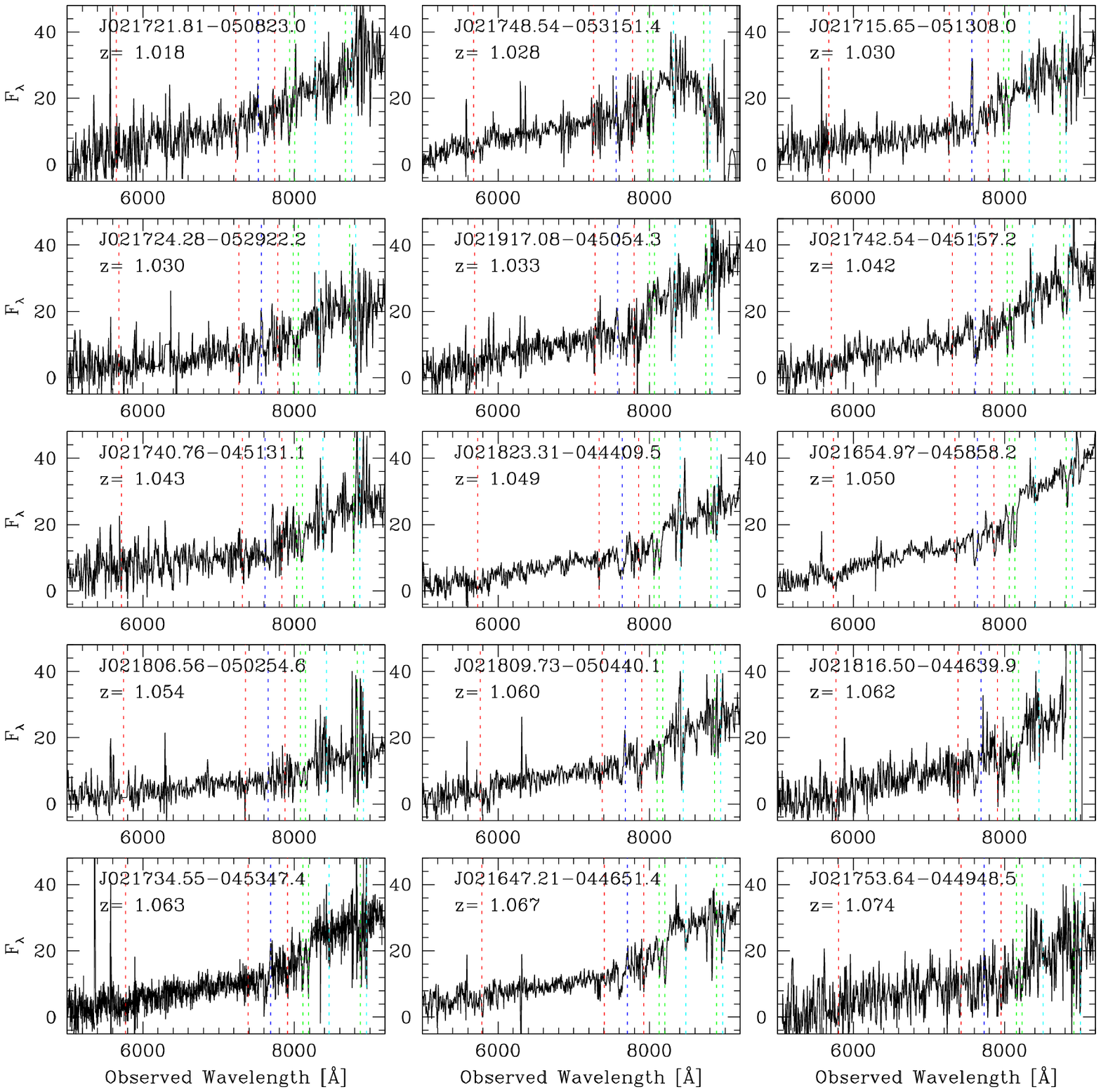}
\caption{Continued.}\label{fig:fig8d}
\end{figure}

\begin{figure}
\figurenum{8}
\plotone{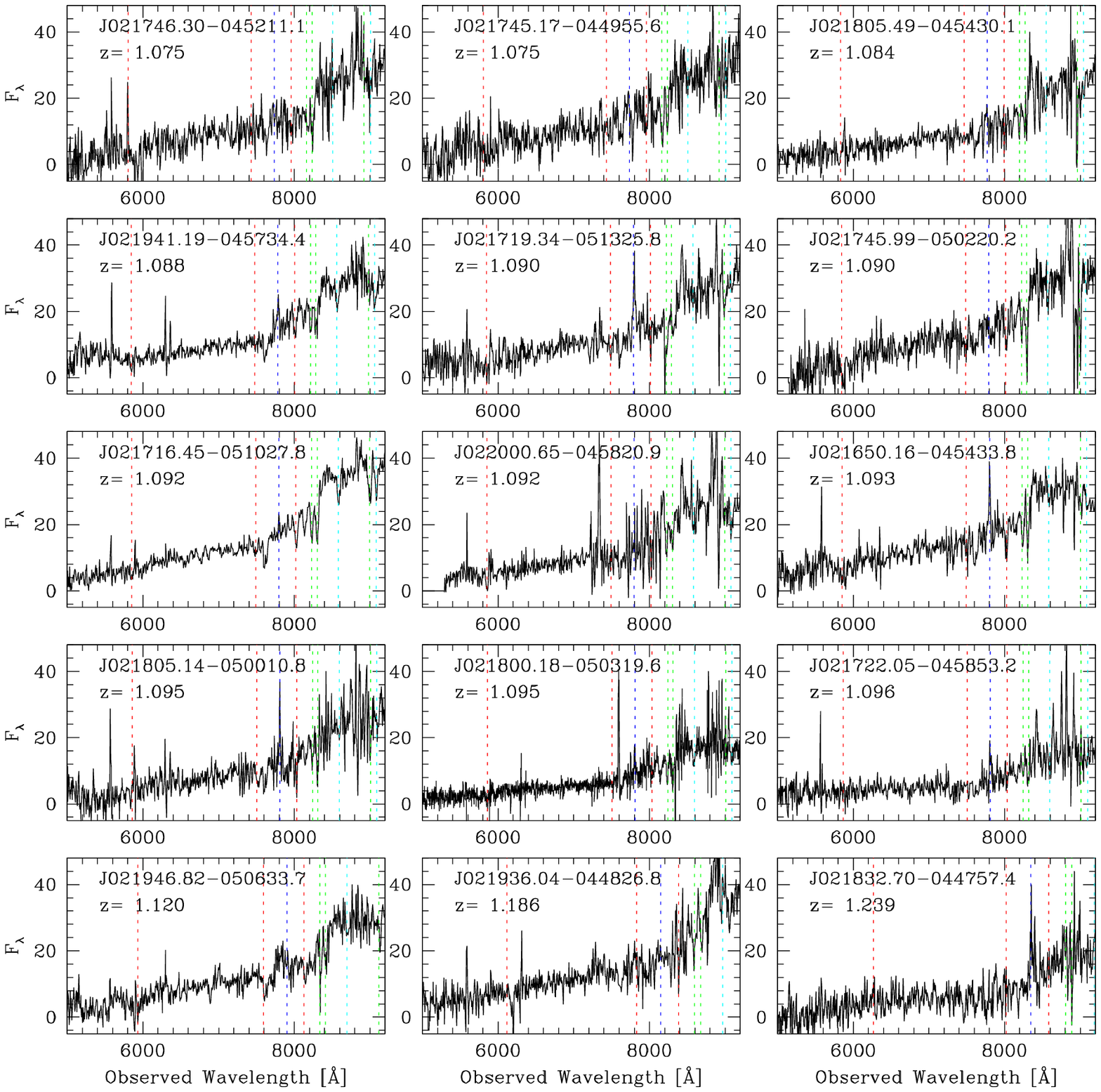}
\caption{Continued.}\label{fig:fig8e}
\end{figure}

\begin{figure}
\figurenum{8}
\plotone{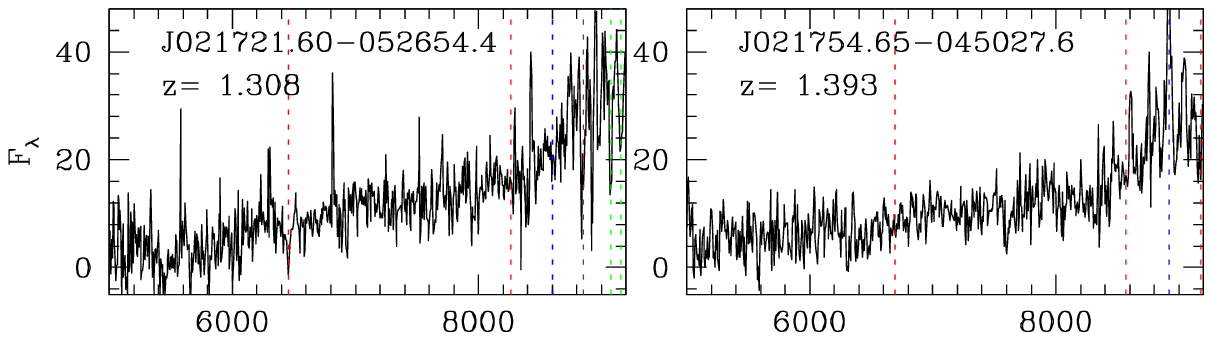}
\caption{Continued.}\label{fig:fig8f}
\end{figure}

\begin{figure}
\figurenum{9}
\plotone{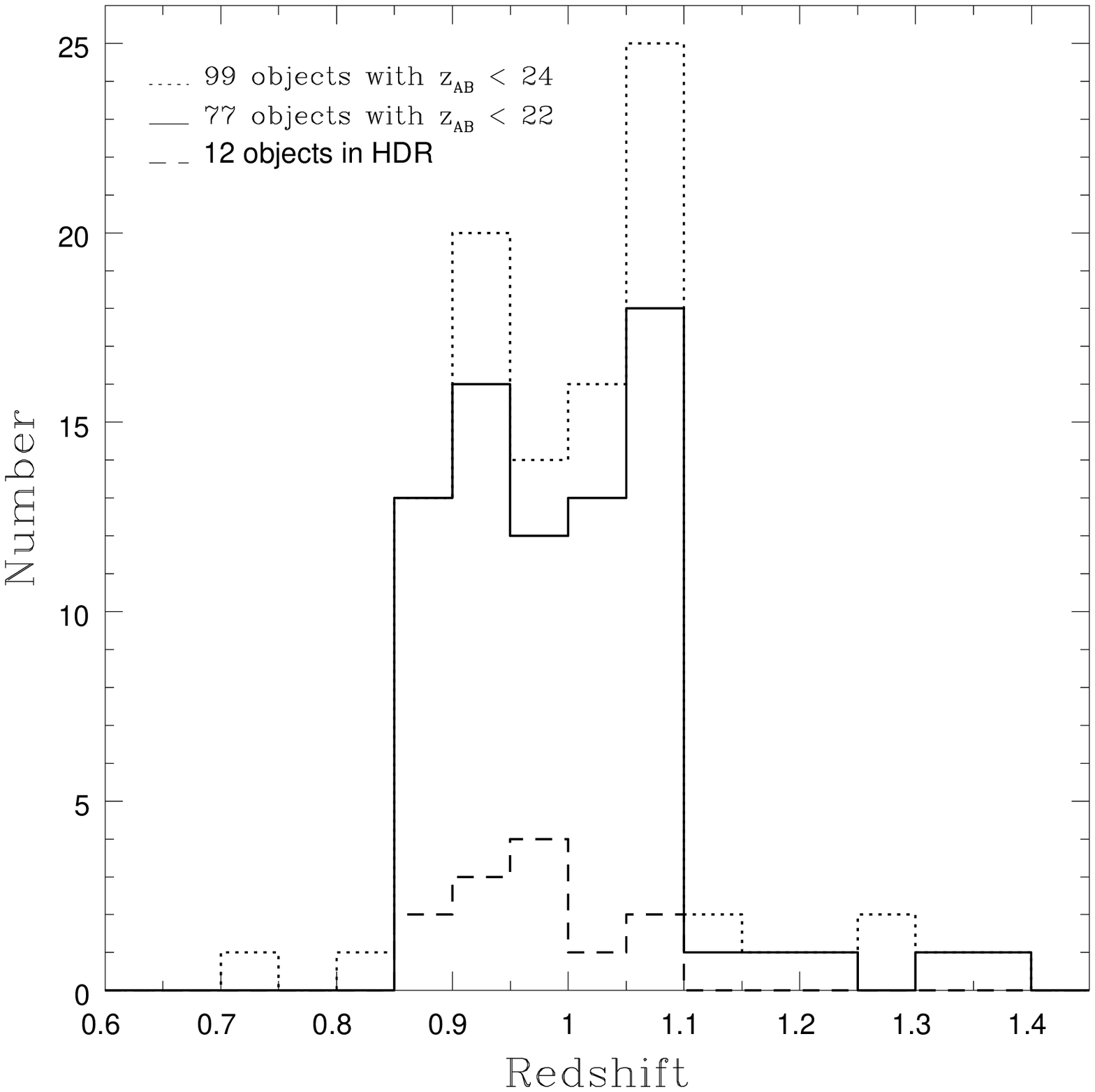}
\caption{The redshift distribution of the spectroscopic sample of OPEGs.}\label{fig:fig9}
\end{figure}

\begin{figure}
\figurenum{10}
\plotone{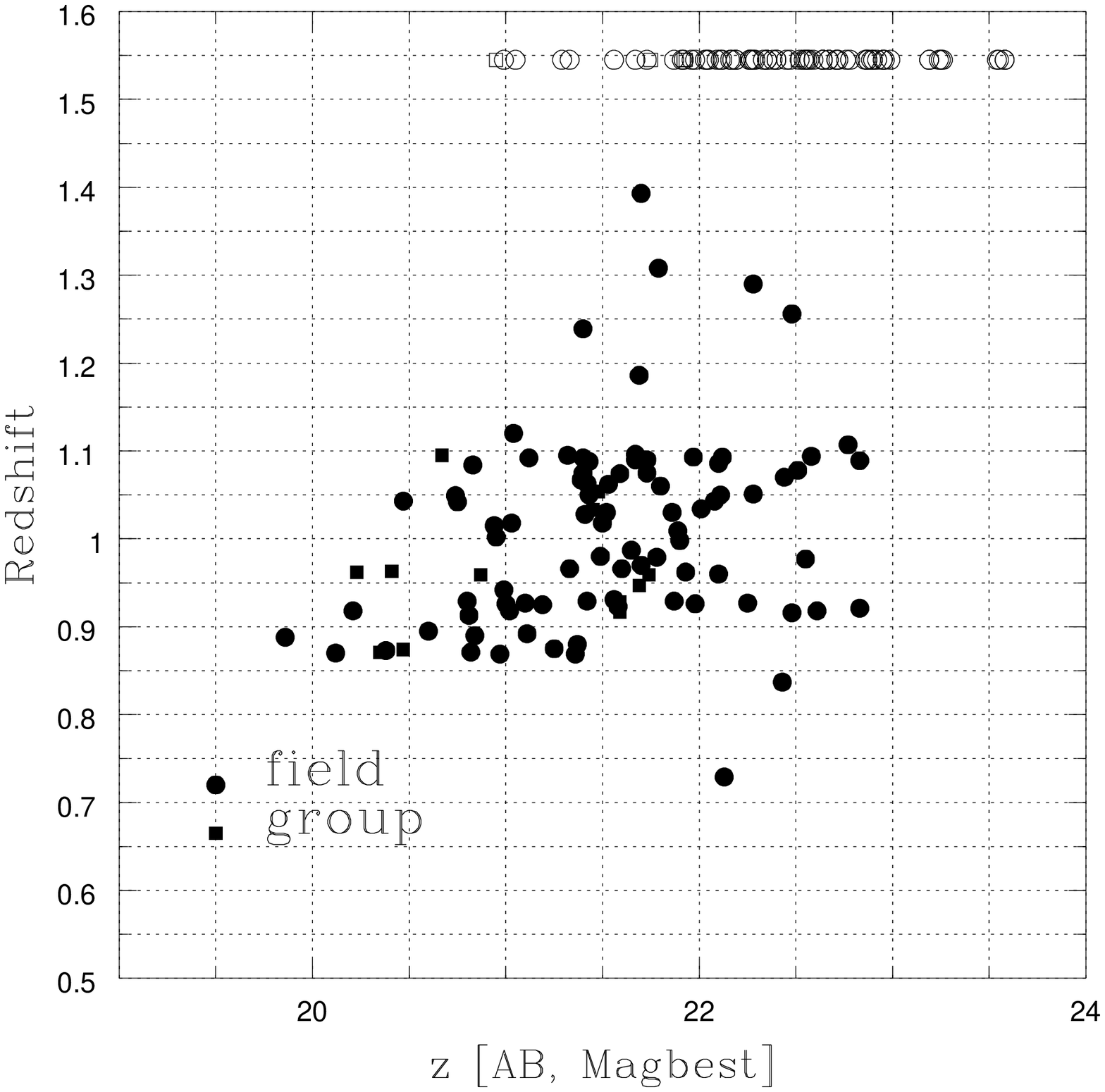}
\caption{The redshift distribution of the OPEG candidates plotted against the $z'$-band magnitude. Magnitude values of the sources whose redshift were not determined are plotted with open circles.}\label{fig:fig10}
\end{figure}

\begin{figure}
\figurenum{11}
\plotone{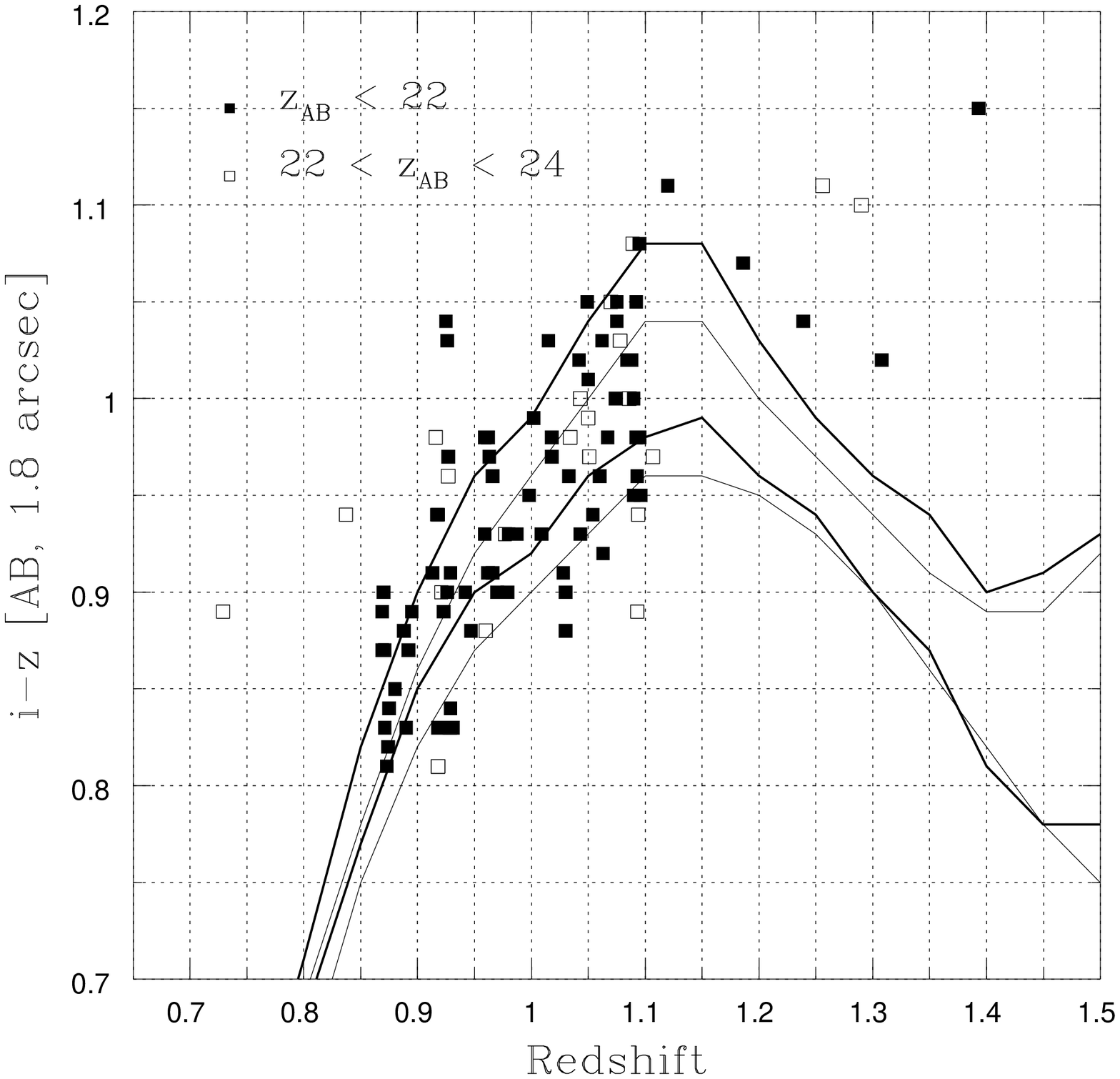}
\caption{$i'-z'$ color and redshift distribution of the OPEG candidates with spectroscopic redshift. Passive evolution models for $z_F$ = 5 (upper) and 2 (lower) and 5$L_*$ (thick) and 0.1$L_*$ (thin). }\label{fig:fig11}
\end{figure}

\section{DISCUSSIONS}

\subsection{Luminosity Function of OPEGs at $z=1$}

 Provided the redshift selection function based on the spectroscopic sample, we discuss the luminosity function of the OPEGs and their spatial density. We here use the simplest procedure to obtain the rest-frame $B$-band luminosity function (LF) of OPEGs at $z=0.9$-1.1 based on the result that the redshift distribution of the $z'<22$ spectroscopic sample show strong peak at the redshift range.

 With minimum assumption, we locate all the selected OPEGs at $z=1$ and adopt the single value of the fraction of galaxies between z=0.9 and 1.1 from the spectroscopic results. The difference in distant modulus between $z=0.9$ and 1.0, and that between $z=1.0$ and 1.1 is 0.28 and 0.26 mag, respectively. The analysis yields magnitude uncertainty of these values, which is still smaller than the bin width of 0.5 mag adopted here. For the spectroscopic sample of 93 $z'<22$ OPEG candidates, 59 are at $z=0.9-1.1$, which yields 63.4$\%$. We simply calculate absolute magnitude at $z=1.0$ for all the OPEG candidates and multiply the factor 0.64 to the density values at {\it all} the magnitude bin. Fig.18 shows the results. We first limit the magnitude range to $M_B=-21$ which corresponds to $z'=22.2$ at $z=1.0$ and is close to the depth of the spectroscopic sample. The red circles are the LF obtained from the raw counts and the blue ones show that after the incompleteness correction using the whole-field data. We discuss the difference from the LF of the local early-type galaxies in the next subsection.

 We then extend the analysis to the true limit of the photometric sample, $z'=25$ as shown in Fig.19 assuming that the redshift selection function and the contamination fraction are the same as the brighter objects above our spectroscopic limit. We note that the true fraction of contamination may increase towards fainter magnitudes due to the background galaxies whose colors match with our criteria. For example, the old red galaxies at $z>1.1$ with some right amount of star-formation could have the similar color as the OPEGS at $z=0.9-1.1$. Thick line is the best-fit Schechter function for the incompleteness-corrected data. Thin lines are the same but for each SXDS-C, -N, -S, -E, and -W fields; density variation is seen as expected from the surface number counts. The resultant Schechter function parameter values are summarized in Table 5.

 For the check, we also calculated the LF from the sample selected by the $3"$-diameter-aperture colors in order to see how large is the uncertainty of the color selection. Indeed, the $i'-z'$ color in $3"$-aperture is $\sim 0.05$ mag bluer than the $2"$-aperture. The results are shown in Fig.20 and Table 5. $3"$-aperture color selection yields slightly but systematically smaller number density and steeper faint-end slope although the difference between the Schechter parameters of the two cases is small and just marginally significant. The difference is due to the fact that the galaxy colors move systematically bluer and a fraction escaped from the model criteria (Fig.4).

\subsection{Comparison with local early-type galaxies}

 The colors of the local counterparts of $z=1$ OPEGs can be inferred by extrapolating the passive models to $z=0$. Although secondary bursts due to minor merger events may change their color in short time scale, it does not affects the spectra of the entire galaxy after a few times 10$^{8}$ yrs (see Sec.2.2.1). Thus the descendants of $z=1$ OPEGs should be among the local galaxies which form the red color-magnitude sequence expected for old ($>10$ Gyr) passive galaxies.

 We here compare the number densities of $z=1$ OPEGs with the red galaxies detected in Slone Digital Sky Survey (SDSS). To avoid any systematic zero-point error, we consider the color range from the observed red sequence of the local galaxies instead of directly extrapolating the models to define the OPEGs at $z=1$. The color-selected $r'$-band LF of the SDSS red-sequence galaxies is kindly provided by Osamu Nakamura (in private communication). We referred the red sequence (zero points) of the morphologically-selected E/S0 galaxies in Fukugita et al. (2004) and adopted the color cut at the blue side $\Delta (g'-r')$=$-0.1$. The $r'$-band LF was calculated as in the similar manner in Nakamura et al. (2003) and is shown in Fig.18 by the thick dashed line; we assume a single representative color, $B-r'$=1.2, to shift it for the $B$-band. 

  It is also interesting to compare $z=1$ OPEGs and the morphologically-selected E/S0 galaxies at $z=0$ themselves since most of the local early-type galaxies have no or only a very small amount of young stellar population and their spectra can be well explained by the passive evolution models with old age (e.g., Kodama and Arimoto 1997). We compare our result with the LF of the morphologically-selected early-type galaxies by Nakamura et al. (2003), which is also plotted in Fig.18 by the solid thick line. The magnitude shift, $B-r'$=1.2, is also adopted. The number density of the color-selected red galaxies in SDSS is in fact larger than that of the morphologically-selected ones. The difference may be due to the presence of the morphologically later-type galaxies with similarly red colors.

 It is clear in Fig.18 that the number density of the OPEGs more luminous than $M_B=-21$ is much larger than those of the local 'early-type' galaxies. 
 
 We plotted in Fig.18 the {\it evolved} luminosity function; pure luminosity evolution for passive models with $z_f=2$ and $z_f=5$ are plotted with the cyan and the green lines, respectively. We used the representative magnitude shift values, $\Delta M_B = -1.5$ and $-1.1$ for the cases of $z_f=2$ and 5, respectively. The prediction with $z_f>3$ is similar ($\Delta M_B < 0.1$ mag) with the case of $z_f=5$. Since our sample of OPEGs is selected by their colors based on the similar passive evolution models, we can directly compare the number densities of the observed results with these pure passive models in a very consistent manner. The number density of OPEGs at $z\sim1$ is $\approx 65$, 50, and 30$\%$ of the predicted ones of the evolved local color-selected red galaxies and $\approx 90$, 85 and 85$\%$ of the local morphologically-selected galaxies at the evolved absolute magnitude $M_B= M_*-1$, $M_*$, $M_*+1$, respectively, for the case of $z_f > 3$.  If the local early-type galaxies are as young as $z_f=2$, OPEGs occupy only $\approx 20$-40$\%$ and $\approx 40$-$80\%$ of the progenitors of the local color-selected and morphologically-selected early-type galaxies, respectively, between $M_B= M_*-1$ and $M_*+1$. As it is not realistic that all the local early-type galaxies are as young as $z_f=2$, as we see at least some evolved galaxies at $z > 2$ (Daddi et al. 2005), it is fair to say that more than half, but not all, of the present-day typical ($M_*-1 < M < M_*+1$) early-type galaxies had already been formed into luminous quiescent galaxies at $z = 1$.

 The results also mean that the less than half, but non-negligible fraction of the present-day early-type galaxies have been evolved since $z \sim 1$. There are at least three possibilities to explain the result; more star-formation at $z \sim 1$, the younger age (they are quiescent but formed at $1 < z < 2$), and the evolution via merging. Some different star-formation history to intermediate redshift may be allowed so that they are observed as OPEGs at $z=0$ if there are enough long time ($\sim 5$ Gyr) after the star-formation ceased (e.g., Bower, Kodama, Terlevich 1999). That is also true for the case that galaxies are younger than $z=2$, i.e., the major star-formation finished between $z=1$ and 2. They are too blue to be classified as the OPEGs defined here but again there is enough time to be the OPEGs at $z=0$. In either case, the progenitors can be among the different category of the objects (e.g., late-type morphology). There is also a possibility that a fraction of the early-type galaxies at $z=0$ have experienced a major merger event as other recent observational results also suggest (e.g., Bell et al. 2005; van Dokkum et al. 2005). Even for the case of the 'dry merger' (Bell et al. 2005), the resultant luminosity shift of factor $\sim 2$ can explain the difference between our $z=1$ OPEG LF and the LF of $z=0$ morphologically-selected early-type galaxies. 

 On the other hand, our results show that stars in more than half of the OPEGs at $z=1$ were formed by $z \sim 2$. It can be tested by observing the OPEG LF between $z=1$ and 2 whether such process as 'dry merger' dominantly occurred or not; if we see much less number density of OPEGs than predicted by the passive evolution models, that is to be evidence that galaxy assembly occurred later than their star formation for the early-type galaxies in general field.

\subsection{Comparison with other results at intermediate redshift}

 We also compare our results with the number densities of old or early-type galaxies at intermediate redshift selected with various different criteria.

 There were many results about the number density of color- or morphologically-selected early-type galaxies at $z \sim 1$ (Lilly et al. 1996; Kauffmann, Charlot, \& White 1996; Zepf 1997; Totani \& Yoshii 1998; Francheschini et al. 1998; Berger et al. 1999; Schade et al. 1999; Menanteau et al. 1999; Treu \& Stiavelli 1999; Thompson et al. 1999; Yan et al. 2000; Kajisawa and Yamada 2001; Im et al. 2002; Daddi et al. 2002; Cimatti et al. 2002a, b, c; Roche et al. 2002; Miyazaki et al. 2003; Pozzetti et al. 2003; Glazebrook et al. 2004; Bell et al. 2004; Saracco et al. 2005; Treu et al. 2005). Treu (2003) present an excellent review for these previous works. The samples of galaxies in these previous study, however, are limited to relatively small area of at most several hundred arcmin$^2$ or to the small numbers of at most a few hundred.

 Miyazaki et al. (2003) obtained the luminosity functions of their sample of Extremely Red Objects. They first selected EROs by the criteria $R-K_{\rm AB} > 3.3$ and then classified them to 'old galaxies' (OGs) and the 'dusty galaxies' by the analysis of the multi-band SED data. They found that the luminosity function of the OGs are consistent with the passive evolution prediction from the LF of the $z=0$ morphologically-selected E/S0 galaxies by Marzke et al. (1998). In fact, our OPEG LF also coincides with the passive-evolution prediction using the local LF of Marzke et al. (1998) but Marzke et al. (1998) gives lower number density compared with Nakamura et al. (2003). In this sense, our results seem consistent with the LF by Miyazaki et al. (2003) but at the same time the conclusion is different and that the LF of early-type galaxies do show evolution if we refer the local LF by Nakamura et al. (2003) based on the homogeneous large sample using the SDSS database.

 Bell et al. (2003) reported their results of the COMBO-17 survey of 0.78 deg$^2$. The multi-band photometric data allow them to evaluate photometric redshift of the galaxies and to make the robust spectral classification in the volume comparable (60$\%$) with SXDS. The limiting magnitude of the sample is $\sim 22$ AB magnitude at 9140 \AA\ , which is roughly equal to an OPEG with $L_*$ luminosity at $z=0$ and 3 magnitude shallower than the sample in this paper. They found the strong color bimodality among their sample and the mean color of the red galaxy sequence evolves with redshift in a way that is consistent with the aging of old stellar population. At the same time, their $B$-band luminosity function of the red-sequence galaxies shows very strong evolution toward $z=1$. The density normalization $\phi_*$ at $z \sim 1$ is almost $15\%$ of the value at $z=0$. In our data, however, no such strong evolution in the density normalization is seen. At $M_B=-21$, the density of the red-sequence galaxies at $z=0.95$ in Bell et al. (2004) is only $\approx 50 \%$ of the OPEGs at $z=1$ in this paper. The difference may not be due to the difference of the color-selection criteria. While Bell et al. first fitted the color-magnitude sequence (based on the photometric redshifts) and then take the galaxies down to 0.25 mag blueward in rest $U-V$ color, our criteria sample the galaxies toward $\approx 0.15$ mag bluer $R-z'$ (close to the rest $U-B$) color from the peak (Fig.4). For the typical galaxy SED, $\Delta (U-V)=0.25$ corresponds to $\Delta (U-B) \approx 0.15$. Sample completeness or field-to-field variation should be examined further to understand the difference.

 Chen et al. (2003) studied the number density of color-selected early-type galaxies using  the $H$-band selected sample of the $\sim 0.35$ deg$^2$ area. Their detection limit is $H=20.8$ (5$\sigma$). Since their color classification of red galaxies include not only E/S0 galaxies but also Sab galaxies, it is difficult to directly compare their result with ours. They obtained a result that the number density of these early-type galaxies decreases $\sim 40$$\%$ since $z=1$. 

 Among these studies, the current sample of SXDS $z=1$ color-selected early-type galaxies, or OPEGs, is the largest in both the sample size (the number of galaxies) and the survey area, which must considerably increase the statistical accuracy of the results. Moreover, we made a pilot spectroscopic observation to calibrate our 'photometric' sample, although the number of the galaxies with redshift information is still limited. 

\section{Summary}

 We have constructed a large sample of the $\approx 4000$ color-selected old passively-evolving galaxies (OPEGs) at $z=0.85-1.1$ using the notably deep and wide-field imaging data of the Subaru/XMM-Newton Deep Survey. Although our sample selection was first made by the simple color-cut based on the galaxy-evolution models but the obtained color and the magnitude distributions as well as the results of the spectroscopic observations of $\approx 80$ galaxies strongly supports that the sample is indeed dominated by the OPEGs at $z=0.85-1.1$. The number-density of the OPEGs still show 10-$30\%$ variations over the 0.5 deg scale and the whole average density is $\approx 40-50\%$ of those of the local color-selected red galaxies and 60-85$\%$ of the local morphologically-selected galaxies at $M_B \sim M_* $. In the near future, deep near-infrared data from the UKIDSS project as well as the mid-infrared data from the SWIRE survey provides further complete SED coverage and the whole evolutionary feature of OPEGs can be studied over large redshift-range.

\vspace{0.5cm}

 We thank the staff of the Subaru Telescope for their assistance with
 our observations. This work is partially supported by the grants-in-aid
 for scientific research of the Ministry of Education, Culture, Sports,
 Science, and Technology (14540234, 15740126). 
 Based in part on data collected at
 Subaru Telescope and obtained from data archive at 
 Astronomical Data Analysis Center, which are operated 
 by the National Astronomical Observatory of Japan.
 The Image Reduction and Analysis
 Facility (IRAF) used in this paper is distributed by National Optical
 Astronomy Observatories. U.S.A., operated by the Association of
 Universities for Research in Astronomy, Inc., under contact to the
 U.S.A. National Science Foundation. 

{}

\newpage
\begin{deluxetable}{lccrrrccc}
\tabletypesize{\small}
\tablecaption{Summary of the SXDS Catalog \label{tbl-1}}
\tablewidth{0pt}
\tablehead{
 \colhead{Field} & \colhead{A$_{\rm clean}$} & \colhead{FWHM}  & \colhead{T$_{\rm exp}$($R$)} & \colhead{T$_{\rm exp}$($i'$)} & \colhead{T$_{\rm exp}$($z'$)} &  \colhead{$R_{\rm lim}$}           & \colhead{$i'_{\rm lim}$} &  \colhead{$z'_{\rm lim}$} \\
 \colhead{ }       & \colhead{arcmin$^2$}  &  \colhead{arcsec} &  \colhead{sec}     & \colhead{sec}         &\colhead{sec}   & \colhead{mag} & \colhead{mag}     & \colhead{mag} \\
}
\startdata
 SXDS Center &  960.1 & 0.80 & 14880 & 38820 & 13020 & 27.0  & 27.1 & 26.0 \\
 SXDS North  &  938.2 & 0.84 & 13920 & 26405 & 15122 & 27.2  & 27.1 & 26.1 \\
 SXDS South  &  955.8 & 0.82 & 13920 & 18540 & 11040 & 27.1  & 26.9 & 25.8 \\
 SXDS East   &  905.3 & 0.82 & 13920 & 22060 & 15990 & 27.0  & 26.9 & 25.9 \\
 SXDS West   &  904.4 & 0.82 & 13920 & 35883 & 18660 & 27.0  & 27.0 & 26.1 \\
\enddata
\end{deluxetable}
Note. -- The limiting magnitude are the 5$\sigma$ values. 
\newpage

\begin{deluxetable}{lccc}
\tabletypesize{\small}
\tablecaption{The Number of the OPEG Candidates \label{tbl-2}}
\tablewidth{0pt}
\tablehead{
 \colhead{Field} & \colhead{A$_{\rm used}$} & \colhead{n($z'<25$)} & 
 \colhead{n$_{\rm OPEG}$($z'<25$)}  \\
 \colhead{ }     & \colhead{arcmin$^2$} & \colhead{}       & \colhead{}  \\
}
\startdata
 SXDS Center &  832.3 &  39166  &  987  \\
 SXDS North  &  811.1 &  39670  &  981  \\
 SXDS South  &  830.2 &  38138  &  653  \\
 SXDS East   &  781.2 &  37201  & 1016  \\
 SXDS West   &  778.3 &  35629  &  843  \\
 SXDS All    & 3690.0 & 175601  &  4118  \\
\enddata
\end{deluxetable}

\newpage

\begin{deluxetable}{ccccc}
\tabletypesize{\small}
\tablecaption{Summary of the Spectroscopic Observations \label{tbl-3}}
\tablewidth{0pt}
\tablehead{
 \colhead{ }   & \colhead{field} & \colhead{ } & 
 \colhead{HDR} & \colhead{ }   \\
 \colhead{ }   & \colhead{n(with z)} & \colhead{n(no z)} &
 \colhead{n(with z)} & \colhead{n(no z) }
}
\startdata
 $19<z'<22$      &  65 &  12   & 12  &  4  \\
 $22<z'<24$      &  22 &  58   &  0  &  2  \\
 total           &  87 &  70   & 12  &  6  \\
\enddata
\end{deluxetable}

\newpage

\begin{deluxetable}{cccccc}
\tablecolumns{6}
\tablenum{4}
\tablewidth{30pc}
\tablecaption{Redshifts of $z'<22$ OPEGs}
\tablehead{
\multicolumn{1}{c}{Name}            &
\multicolumn{1}{c}{$z_{\rm best}$}        &
\multicolumn{1}{c}{$i'-z'$}        &
\multicolumn{1}{c}{$R-z'$}        &
\multicolumn{1}{c}{Redshift$^a$} &
\multicolumn{1}{c}{$EW_{obs}(OII)$} \\
\colhead{} &
\colhead{(mag)} &
\colhead{(mag)} &
\colhead{(mag)} &
\colhead{} &
\colhead{\AA } \\
}

\startdata
J021917.79$-$050442.7 & 21.36 &  0.87 &  1.85 & 0.869&   0 \nl 
J021931.15$-$051240.9 & 20.97 &  0.89 &  1.89 & 0.869&   0 \nl 
J021932.14$-$051257.6 & 20.12 &  0.90 &  1.92 & 0.870&   0 \nl 
J021808.80$-$045831.6 & 20.35 &  0.83 &  1.90 & 0.871&   0 \nl 
J021834.80$-$050140.9 & 20.82 &  0.87 &  1.89 & 0.871&   4 \nl 
J021832.02$-$050059.5 & 20.38 &  0.81 &  1.76 & 0.873&   0 \nl 
J021804.85$-$045913.5 & 20.47 &  0.82 &  1.83 & 0.874&  11 \nl 
J021809.25$-$045449.3 & 21.25 &  0.84 &  1.84 & 0.875:&  0 \nl
J021824.28$-$044640.4 & 21.37 &  0.85 &  1.84 & 0.880&   0 \nl 
J021947.02$-$050823.6 & 19.86 &  0.88 &  1.87 & 0.888&   0 \nl 
J021929.74$-$050544.3 & 20.84 &  0.83 &  1.77 & 0.890&   0 \nl 
J021929.91$-$050602.1 & 21.11 &  0.87 &  1.86 & 0.892&   0 \nl 
J021656.33$-$045521.6 & 20.60 &  0.89 &  1.89 & 0.895&   0 \nl 
J022016.05$-$045521.3 & 20.81 &  0.91 &  1.87 & 0.913&   0 \nl 
J021809.05$-$050160.0 & 21.59 &  0.94 &  1.95 & 0.917&   0 \nl 
J021724.40$-$051251.8 & 20.21 &  0.94 &  1.91 & 0.918&   5 \nl 
J021725.99$-$051119.5 & 21.02 &  0.83 &  1.83 & 0.918&  11 \nl 
J021710.51$-$051333.5 & 21.58 &  0.89 &  1.89 & 0.923&   2 \nl 
J021907.20$-$043150.7 & 21.19 &  1.04 &  1.98 & 0.925&   0 \nl 
J021649.08$-$045509.8 & 21.98 &  0.90 &  1.87 & 0.926&   4 \nl 
J021905.91$-$043141.6 & 21.00 &  1.03 &  1.97 & 0.926&   0 \nl 
J022014.04$-$045025.8 & 21.10 &  0.97 &  1.94 & 0.927&   0 \nl 
J021815.83$-$050306.0 & 21.59 &  0.83 &  1.78 & 0.928&   0 \nl 
J021653.58$-$045632.2 & 21.42 &  0.91 &  1.87 & 0.929&   0 \nl 
J021718.48$-$051027.7 & 20.80 &  0.84 &  1.84 & 0.929&   0 \nl 
J021724.41$-$050508.2 & 21.87 &  0.83 &  1.78 & 0.929&   4 \nl 
J021656.73$-$045719.2 & 21.56 &  0.83 &  1.74 & 0.931&   0 \nl 
J021721.98$-$051156.0 & 20.99 &  0.90 &  1.91 & 0.942&   7 \nl 
J021927.36$-$045111.5 & 21.69 &  0.88 &  1.80 & 0.947&   0 \nl 
J021803.46$-$045852.0 & 21.74 &  0.93 &  1.87 & 0.959&   0 \nl 
J021807.36$-$045925.4 & 20.87 &  0.98 &  1.95 & 0.959&   0 \nl 
J021928.58$-$045040.8 & 20.23 &  0.98 &  1.92 & 0.962&   0 \nl 
J021945.25$-$050353.3 & 21.93 &  0.91 &  1.80 & 0.962&   0 \nl 
J021919.56$-$045119.0 & 20.41 &  0.97 &  1.92 & 0.963&   0 \nl 
J021931.30$-$050631.2 & 21.33 &  0.91 &  1.84 & 0.966&   0 \nl 
J021944.88$-$050521.9 & 21.60 &  0.96 &  1.87 & 0.966&   0 \nl 
J022010.69$-$045456.1 & 21.70 &  0.90 &  1.78 & 0.970&   0 \nl 
J021940.13$-$050746.8 & 21.78 &  0.90 &  1.81 & 0.979&   0 \nl 
J021844.94$-$050303.9 & 21.49 &  0.93 &  1.89 & 0.980&   0 \nl 
J021736.51$-$051035.9 & 21.65 &  0.93 &  1.92 & 0.987&   5 \nl 
J021719.15$-$051447.7 & 21.90 &  0.95 &  1.91 & 0.998&   0 \nl 
J022025.40$-$050346.9 & 20.95 &  0.99 &  1.94 & 1.002&   0 \nl 
J021718.73$-$050932.4 & 21.89 &  0.93 &  1.85 & 1.009&   3 \nl 
J021708.79$-$045731.7 & 20.94 &  1.03 &  1.99 & 1.015&   0 \nl 
J021704.04$-$044753.6 & 21.03 &  0.97 &  1.86 & 1.018&   0 \nl 
J021721.81$-$050823.0 & 21.50 &  0.98 &  1.91 & 1.018:&   0 \nl
J021748.54$-$053151.4 & 21.41 &  0.91 &  1.79 & 1.028&   0 \nl 
J021715.65$-$051308.0 & 21.52 &  0.90 &  1.88 & 1.030&  17 \nl 
J021724.28$-$052922.2 & 21.86 &  0.88 &  1.86 & 1.030&  28 \nl 
J021917.08$-$045054.3 & 21.45 &  0.96 &  1.84 & 1.033:&  15 \nl
J021742.54$-$045157.2 & 20.75 &  1.02 &  1.93 & 1.042&   0 \nl 
J021740.76$-$045131.1 & 20.47 &  0.93 &  1.91 & 1.043&   4 \nl 
J021823.31$-$044409.5 & 20.74 &  1.05 &  1.93 & 1.049&   0 \nl 
J021654.97$-$045858.2 & 21.43 &  1.01 &  1.96 & 1.050&   0 \nl 
J021806.56$-$050254.6 & 21.48 &  0.94 &  1.78 & 1.054&   4 \nl 
J021809.73$-$050440.1 & 21.80 &  0.96 &  1.85 & 1.060&   6 \nl 
J021816.50$-$044639.9 & 21.53 &  1.03 &  1.89 & 1.062&   0 \nl 
J021734.56$-$045347.4 & 21.42 &  0.92 &  2.02 & 1.063&  10 \nl 
J021647.21$-$044651.4 & 21.39 &  0.98 &  1.86 & 1.067&   0 \nl 
J021753.64$-$044948.5 & 21.59 &  1.00 &  1.87 & 1.074&  10 \nl 
J021745.18$-$044955.6 & 21.73 &  1.05 &  1.93 & 1.075&   0 \nl 
J021746.30$-$045211.1 & 21.40 &  1.04 &  1.91 & 1.075&   0 \nl 
J021805.49$-$045430.1 & 20.83 &  1.02 &  1.93 & 1.084&  10 \nl 
J021941.19$-$045734.4 & 21.43 &  1.02 &  1.93 & 1.088&   2 \nl 
J021719.34$-$051325.8 & 21.67 &  1.00 &  1.94 & 1.090&   0 \nl 
J021745.99$-$050220.2 & 21.73 &  0.95 &  1.81 & 1.090&  27 \nl 
J021716.45$-$051027.8 & 21.12 &  1.05 &  1.95 & 1.092&   2 \nl 
J022000.65$-$045820.9 & 21.40 &  0.98 &  1.84 & 1.092&   0 \nl 
J021650.16$-$045433.8 & 21.97 &  0.96 &  1.81 & 1.093&  14 \nl 
J021800.18$-$050319.6 & 21.32 &  0.98 &  1.84 & 1.095&   2 \nl 
J021805.14$-$050010.8 & 20.67 &  1.08 &  2.08 & 1.095&  33 \nl 
J021722.05$-$045853.2 & 21.67 &  0.95 &  1.88 & 1.096&  19 \nl 
J021946.82$-$050633.7 & 21.04 &  1.11 &  1.93 & 1.120&   0 \nl 
J021936.04$-$044826.8 & 21.69 &  1.07 &  1.79 & 1.186&   0 \nl 
J021832.70$-$044757.4 & 21.40 &  1.04 &  2.01 & 1.239&  29 \nl 
J021721.60$-$052654.4 & 21.79 &  1.02 &  1.80 & 1.308&   0 \nl 
J021754.65$-$045027.6 & 21.70 &  1.15 &  1.93 & 1.393&  23 \nl 
\enddata
\end{deluxetable}
(a) Redshift values with ':' have uncertainty due to the less prominent spectral features.

\clearpage

\newpage

\begin{deluxetable}{lcccccc}
\tabletypesize{\small}
\tablenum{5}
\tablecaption{Schechter fitting parameters for the B-band Luminosity Function of OPEGs \label{tbl-4}}
\tablewidth{0pt}
\tablehead{ \colhead{ } & \colhead{$M_*$} & \colhead{err.} & 
 \colhead{$\phi$$_*$} & \colhead{err. }  & \colhead{$\alpha$} & \colhead{err.}\\
 \colhead{ }   & \colhead{AB mag.} & \colhead{} &
 \colhead{$\times 10^{-3}$ $h_{70}^3$ Mpc$^{-3}$} & \colhead{ } &\colhead{ } & \colhead{}
}
\startdata
 All (single-z method) &  -21.38 & 0.10 & 1.46 & 0.14 &  -0.67 &  0.07 \\
 Center                &  -21.34 & 0.21 & 1.52 & 0.32 &  -0.72 &  0.15 \\
 North                 &  -21.29 & 0.21 & 1.74 & 0.30 &  -0.60 &  0.15 \\
 South                 &  -21.33 & 0.27 & 1.02 & 0.24 &  -0.63 &  0.18 \\
 East                  &  -21.60 & 0.21 & 1.46 & 0.28 &  -0.76 &  0.14 \\
 West                  &  -21.39 & 0.22 & 1.46 & 0.32 &  -0.69 &  0.16 \\
 All (3$^{\prime\prime}$ aperture) &  -21.25 & 0.11 & 1.38 & 0.12& -0.54 & 0.09  \\
\enddata
\end{deluxetable}

\clearpage

\begin{figure}
\figurenum{12}
\plotone{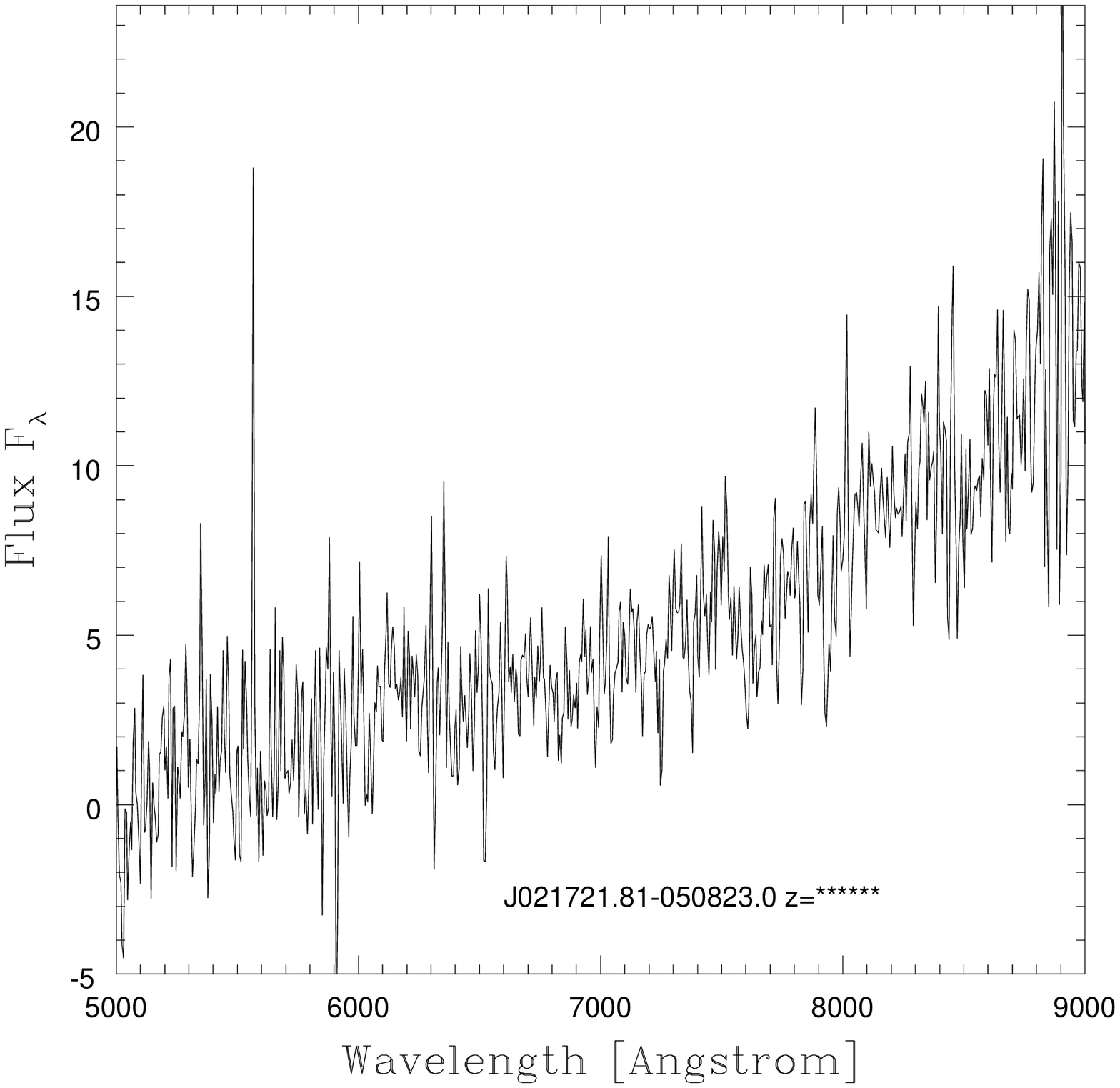}
\caption{An example of the objects with featureless red continuum.}\label{fig:fig12}
\end{figure}

\begin{figure}
\figurenum{13}
\plotone{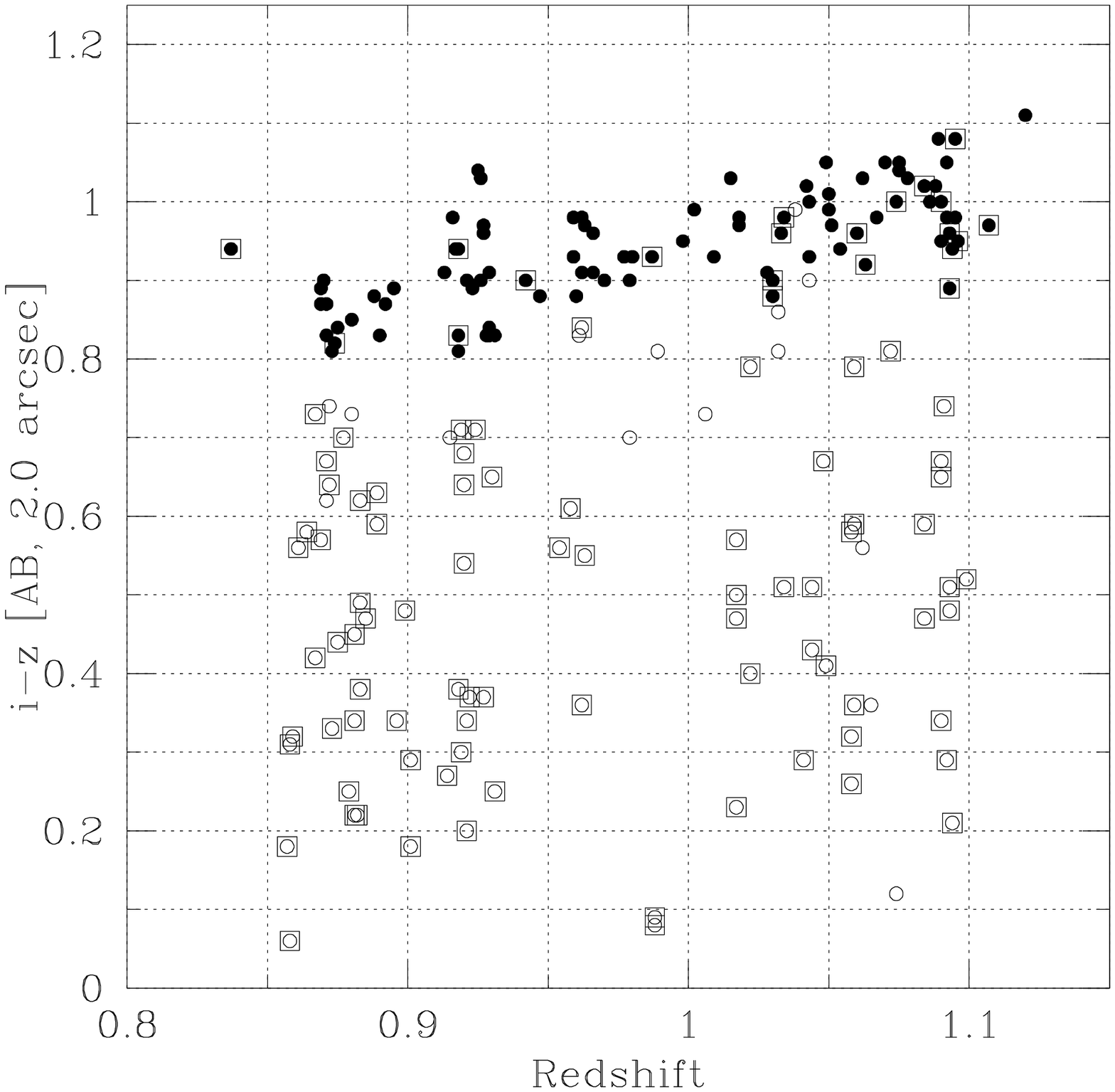}
\caption{Color distribution of the all the galaxies at $0.8 < z < 1.15$. The filled circles are the OPEGs. The open squares show the objects which have the [OII] emission lines with the observed equivalent width $EW_{obs}$(OII)$> 5$ \AA .}\label{fig:fig13}
\end{figure}

\begin{figure}
\figurenum{14}
\plotone{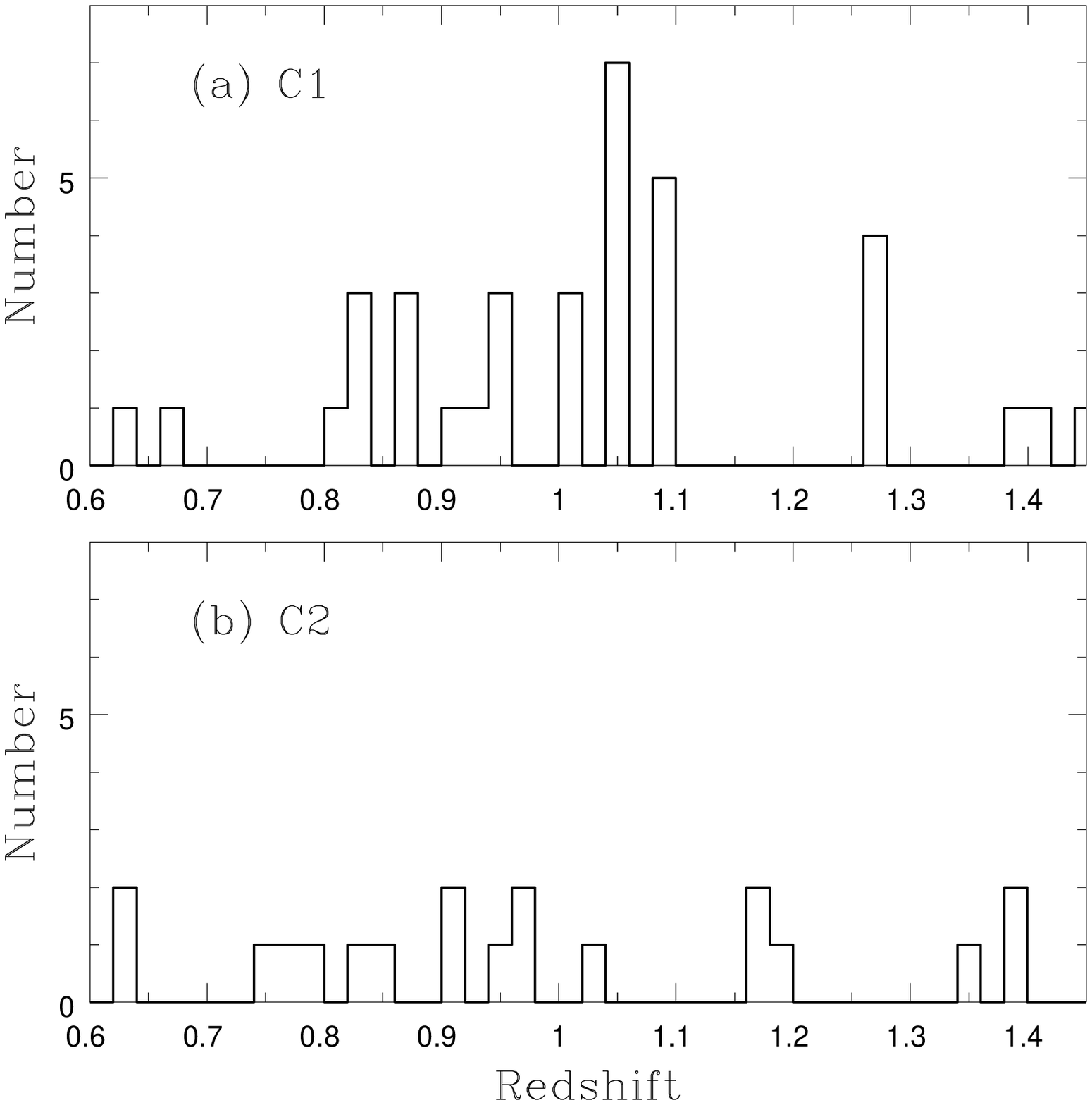}
\caption{Redshift distribution of the galaxies in the 'high density region' no.1 and 2 in Kodama et al. (2004).}{fig:fig14a}
\end{figure}

\begin{figure}
\figurenum{15}
\plotone{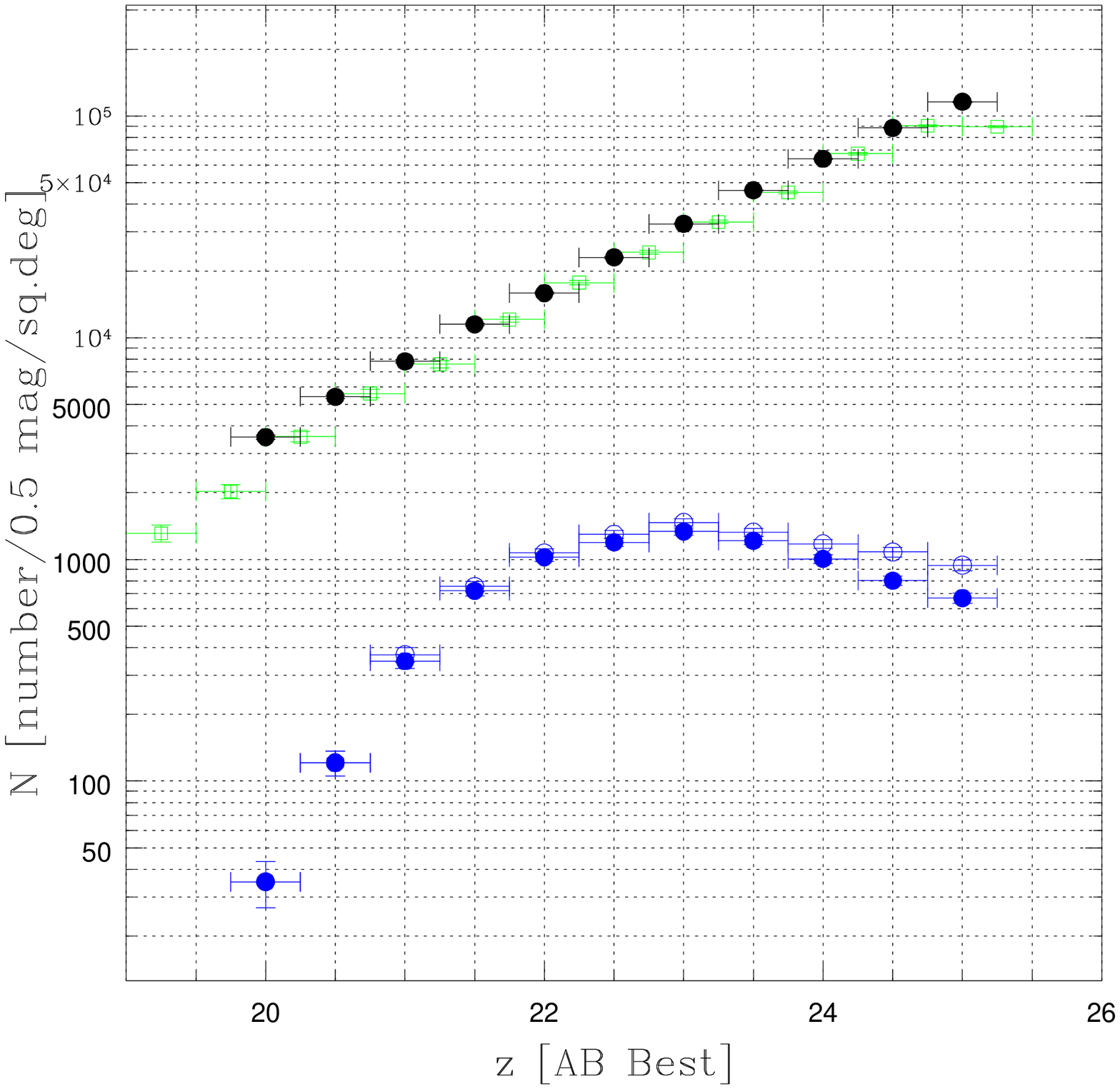}
\caption{Number counts of the $z'$-selected objects (top) and the OPEG candidates (bottom). The values after incompleteness correction for the OPEG candidates are shown by the open circles. The $z'$-band number counts obtained by Capak et al. (2003) using slightly different method is also plotted by the open squares for comparison.}\label{fig:fig15}
\end{figure}

\begin{figure}
\figurenum{16}
\plotone{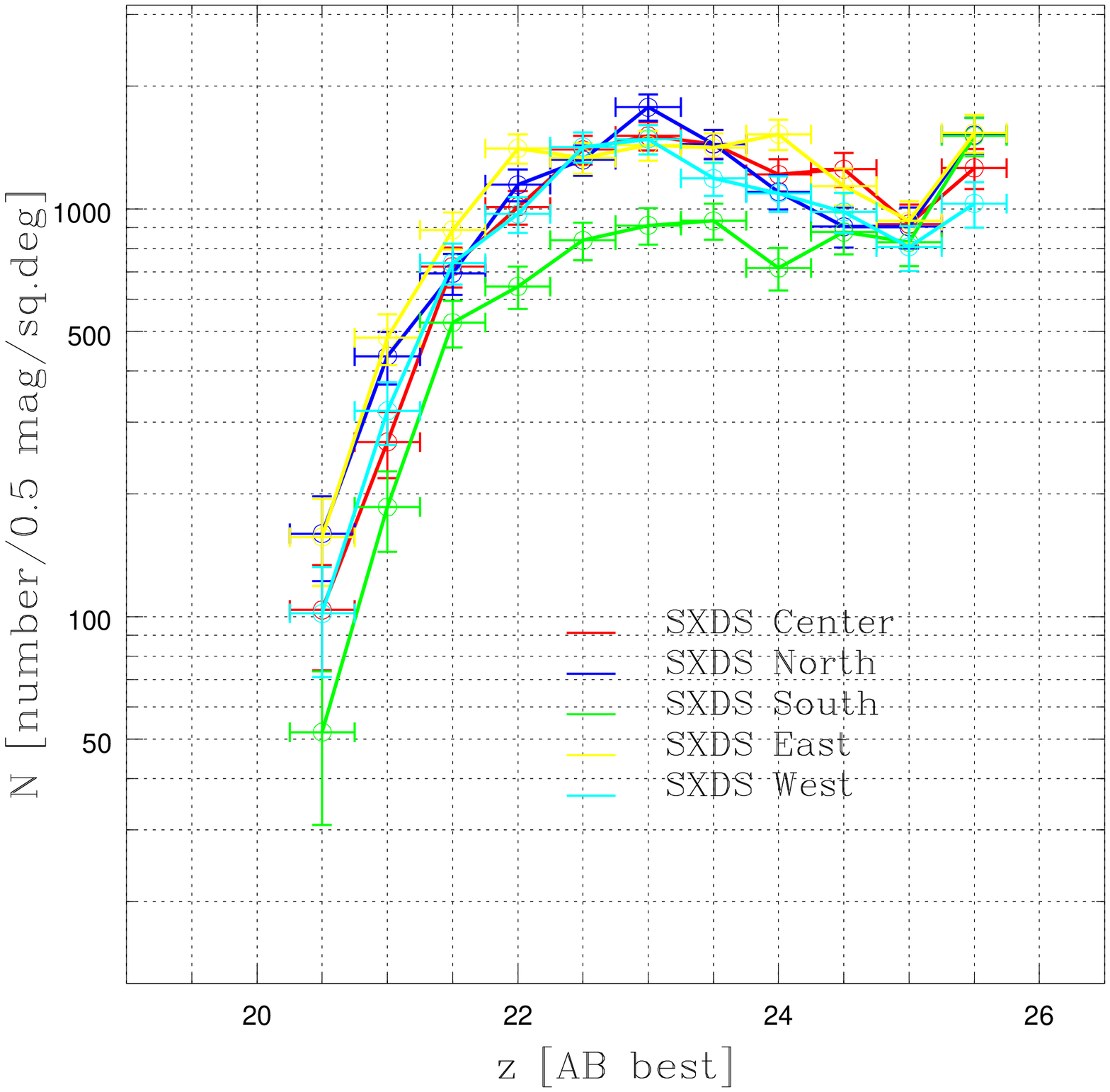}
\caption{Number counts for OPEGs in SXDS-C, -N, -S, -E, -W fields. }\label{fig:fig16}
\end{figure}

\begin{figure}
\figurenum{17}
\plotone{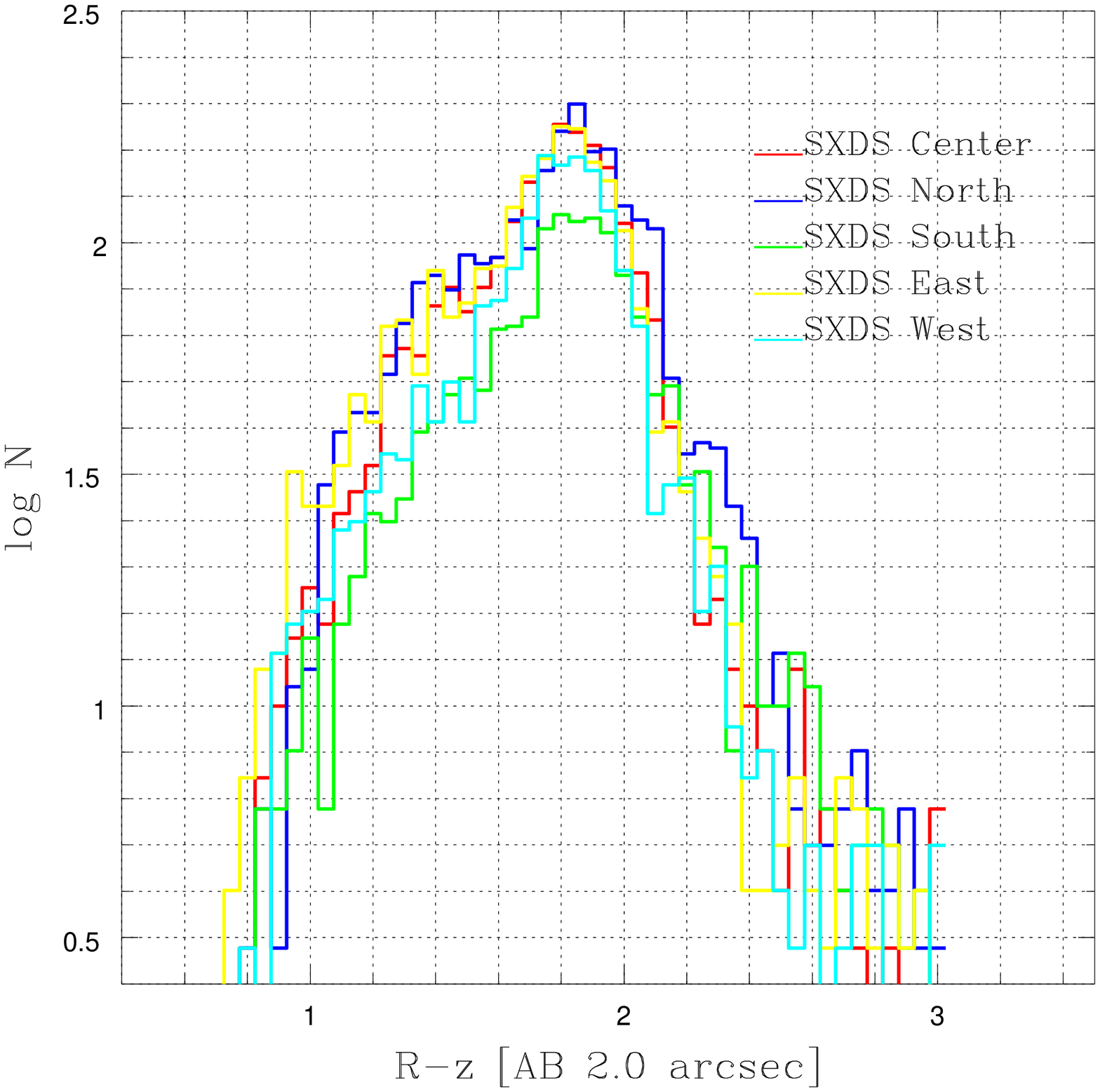}
\caption{$R-z'$ color distribution for the OPEGs in SXDS-C, -N, -S, -E, -W fields. }\label{fig:fig17}
\end{figure}

\begin{figure}
\figurenum{18}
\plotone{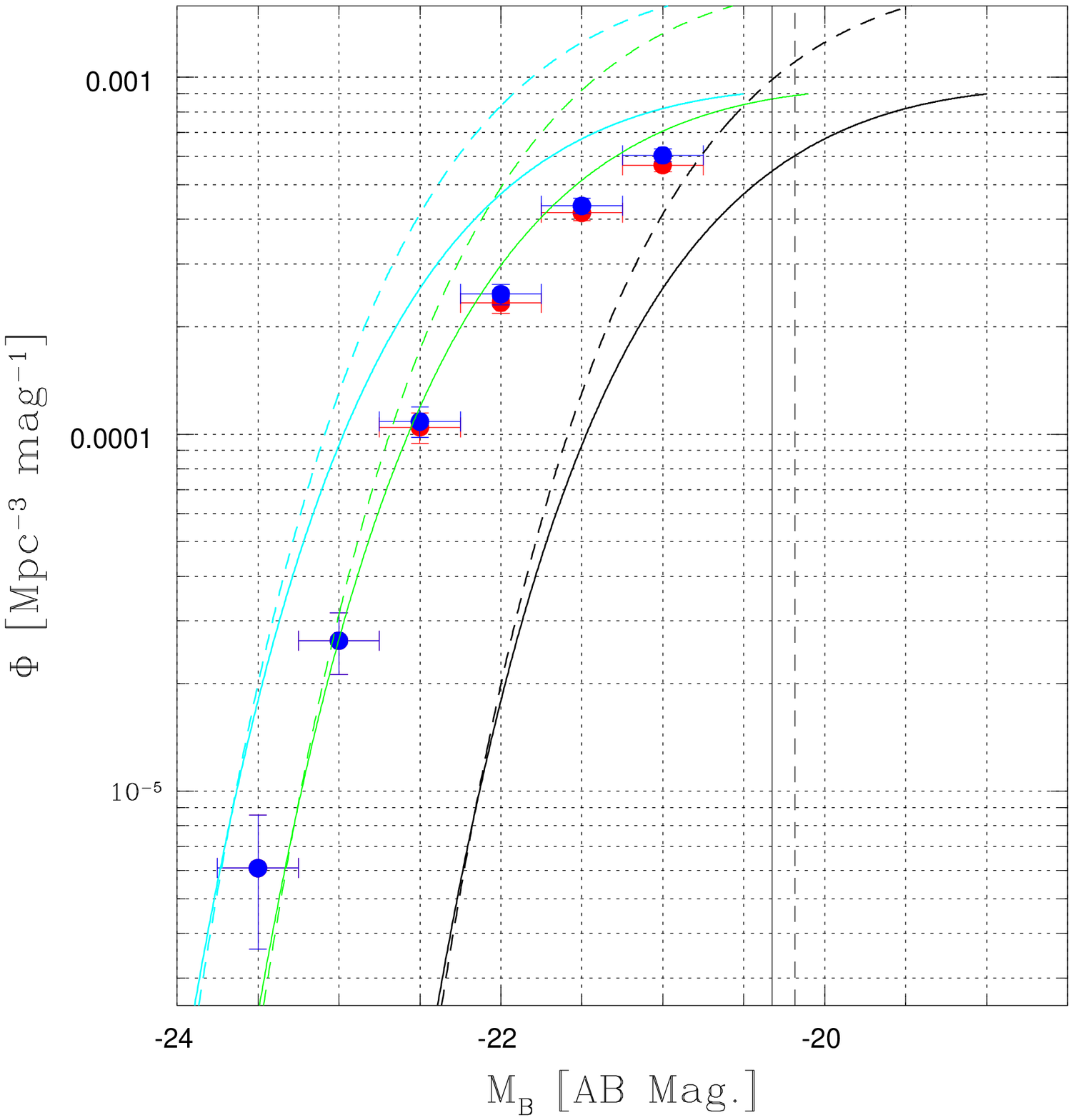}
\caption{$z=1$ OPEG luminosity function at $M_B < -21$ is shown. The red and blue circles show those before and after the correction for the detection incompleteness. For comparison, the local luminosity functions for the SDSS morphologically-selected E/S0 galaxies (solid black line, Nakamura et al. 2003) and for the color-selected red galaxies (dashed black line), as well as those at $z=1$ as expected by pure luminosity evolution with the same passive-evolution models with $z_F$=2 (blue lines) and 5 (green lines) are plotted. The vertical lines show the $M_*$ values for the $z=0$ E/S0 galaxies.}\label{fig:fig18}
\end{figure}

\begin{figure}
\figurenum{19}
\plotone{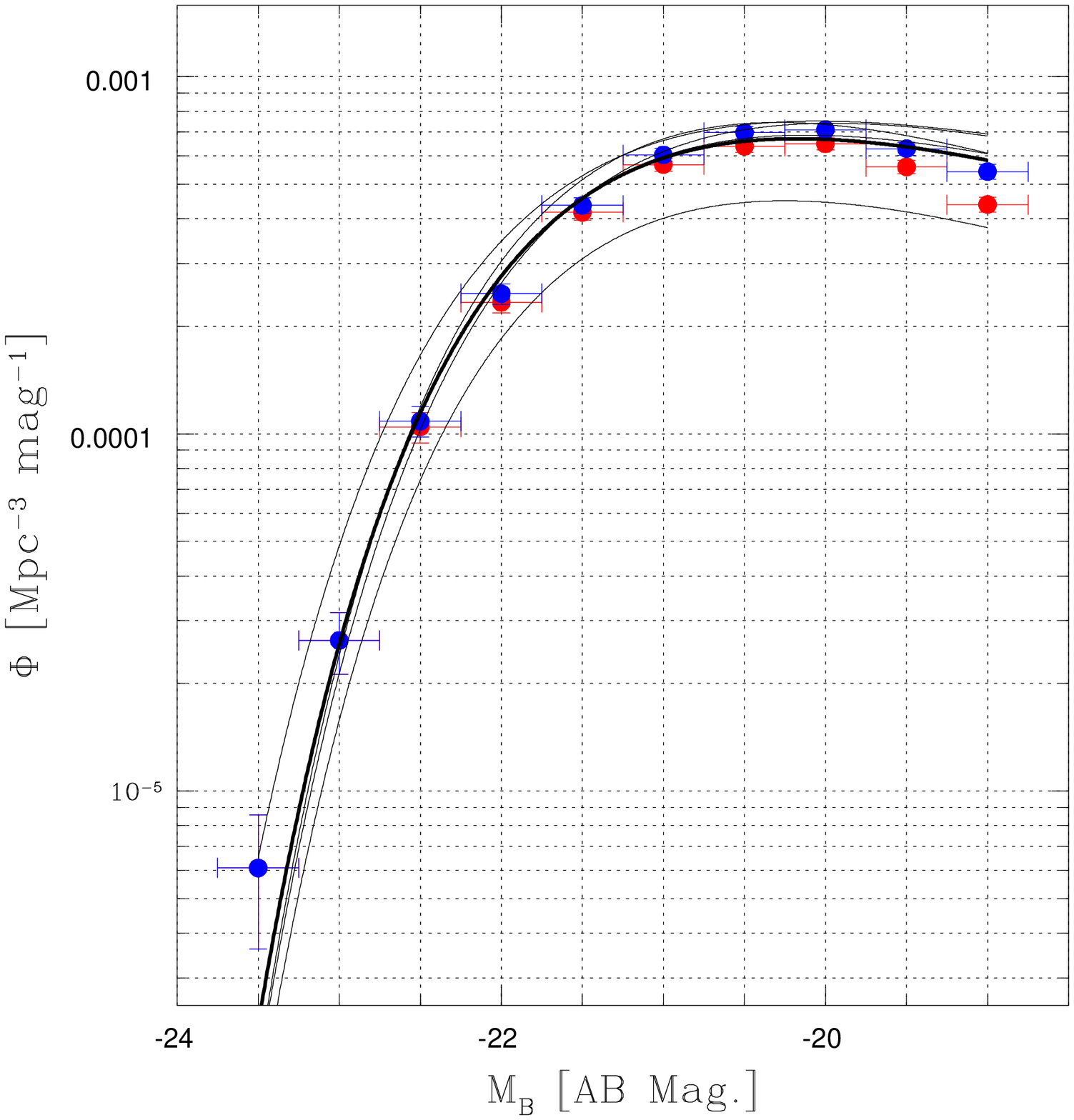}
\caption{$z=1$ OPEG luminosity function. The open (red in the on-line edition) and the filled (blue) circles show those before and after the correction for the detection incompleteness.The Schechter function fit for those obtained in each SXDS-C, -N, -S, -E, -W fields (thin lines) are plotted together.}\label{fig:fig19}
\end{figure}

\begin{figure}
\figurenum{20}
\plotone{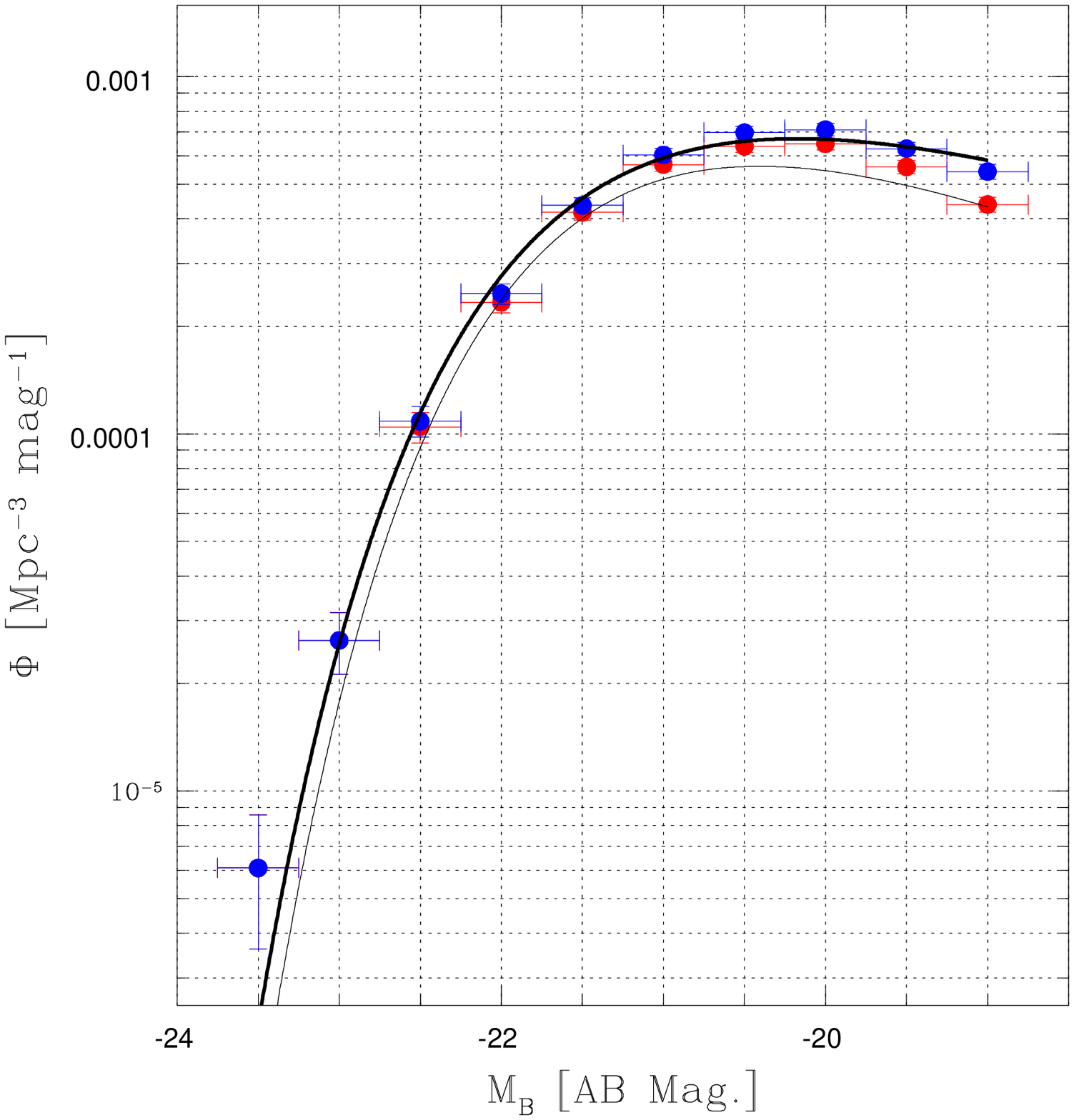}
\caption{$z=1$ OPEG luminosity function (smae as in Fig.19). The Schechter function fit for the $3^{\prime\prime}$-aperture case (thin line) is plotted together.}\label{fig:fig19}
\end{figure}

\end{document}